%% file: main.tex
\documentclass[%
    reprint,
    superscriptaddress,
    amsmath,amssymb,
    aps,
    prl,
]{revtex4-1}

\usepackage{graphicx}
\usepackage{dcolumn}
\usepackage{bm}
\usepackage{color}
\usepackage{multirow}

\newcommand{\beginsupplement}{%
        \setcounter{table}{0}
        \renewcommand{\thetable}{S\arabic{table}}%
        \setcounter{figure}{0}
        \renewcommand{\thefigure}{S\arabic{figure}}%
     }

\graphicspath{{./}{./figures/}}

\begin{document}


\title{Measurement of Atmospheric Neutrino Oscillations at 6-56~GeV with IceCube DeepCore}

\affiliation{III. Physikalisches Institut, RWTH Aachen University, D-52056 Aachen, Germany}
\affiliation{Department of Physics, University of Adelaide, Adelaide, 5005, Australia}
\affiliation{Dept.~of Physics and Astronomy, University of Alaska Anchorage, 3211 Providence Dr., Anchorage, AK 99508, USA}
\affiliation{Dept.~of Physics, University of Texas at Arlington, 502 Yates St., Science Hall Rm 108, Box 19059, Arlington, TX 76019, USA}
\affiliation{CTSPS, Clark-Atlanta University, Atlanta, GA 30314, USA}
\affiliation{School of Physics and Center for Relativistic Astrophysics, Georgia Institute of Technology, Atlanta, GA 30332, USA}
\affiliation{Dept.~of Physics, Southern University, Baton Rouge, LA 70813, USA}
\affiliation{Dept.~of Physics, University of California, Berkeley, CA 94720, USA}
\affiliation{Lawrence Berkeley National Laboratory, Berkeley, CA 94720, USA}
\affiliation{Institut f\"ur Physik, Humboldt-Universit\"at zu Berlin, D-12489 Berlin, Germany}
\affiliation{Fakult\"at f\"ur Physik \& Astronomie, Ruhr-Universit\"at Bochum, D-44780 Bochum, Germany}
\affiliation{Universit\'e Libre de Bruxelles, Science Faculty CP230, B-1050 Brussels, Belgium}
\affiliation{Vrije Universiteit Brussel (VUB), Dienst ELEM, B-1050 Brussels, Belgium}
\affiliation{Dept.~of Physics, Massachusetts Institute of Technology, Cambridge, MA 02139, USA}
\affiliation{Dept. of Physics and Institute for Global Prominent Research, Chiba University, Chiba 263-8522, Japan}
\affiliation{Dept.~of Physics and Astronomy, University of Canterbury, Private Bag 4800, Christchurch, New Zealand}
\affiliation{Dept.~of Physics, University of Maryland, College Park, MD 20742, USA}
\affiliation{Dept.~of Physics and Center for Cosmology and Astro-Particle Physics, Ohio State University, Columbus, OH 43210, USA}
\affiliation{Dept.~of Astronomy, Ohio State University, Columbus, OH 43210, USA}
\affiliation{Niels Bohr Institute, University of Copenhagen, DK-2100 Copenhagen, Denmark}
\affiliation{Dept.~of Physics, TU Dortmund University, D-44221 Dortmund, Germany}
\affiliation{Dept.~of Physics and Astronomy, Michigan State University, East Lansing, MI 48824, USA}
\affiliation{Dept.~of Physics, University of Alberta, Edmonton, Alberta, Canada T6G 2E1}
\affiliation{Erlangen Centre for Astroparticle Physics, Friedrich-Alexander-Universit\"at Erlangen-N\"urnberg, D-91058 Erlangen, Germany}
\affiliation{D\'epartement de physique nucl\'eaire et corpusculaire, Universit\'e de Gen\`eve, CH-1211 Gen\`eve, Switzerland}
\affiliation{Dept.~of Physics and Astronomy, University of Gent, B-9000 Gent, Belgium}
\affiliation{Dept.~of Physics and Astronomy, University of California, Irvine, CA 92697, USA}
\affiliation{Dept.~of Physics and Astronomy, University of Kansas, Lawrence, KS 66045, USA}
\affiliation{SNOLAB, 1039 Regional Road 24, Creighton Mine 9, Lively, ON, Canada P3Y 1N2}
\affiliation{Dept.~of Astronomy, University of Wisconsin, Madison, WI 53706, USA}
\affiliation{Dept.~of Physics and Wisconsin IceCube Particle Astrophysics Center, University of Wisconsin, Madison, WI 53706, USA}
\affiliation{Institute of Physics, University of Mainz, Staudinger Weg 7, D-55099 Mainz, Germany}
\affiliation{Department of Physics, Marquette University, Milwaukee, WI, 53201, USA}
\affiliation{Universit\'e de Mons, 7000 Mons, Belgium}
\affiliation{Physik-department, Technische Universit\"at M\"unchen, D-85748 Garching, Germany}
\affiliation{Institut f\"ur Kernphysik, Westf\"alische Wilhelms-Universit\"at M\"unster, D-48149 M\"unster, Germany}
\affiliation{Bartol Research Institute and Dept.~of Physics and Astronomy, University of Delaware, Newark, DE 19716, USA}
\affiliation{Dept.~of Physics, Yale University, New Haven, CT 06520, USA}
\affiliation{Dept.~of Physics, University of Oxford, 1 Keble Road, Oxford OX1 3NP, UK}
\affiliation{Dept.~of Physics, Drexel University, 3141 Chestnut Street, Philadelphia, PA 19104, USA}
\affiliation{Physics Department, South Dakota School of Mines and Technology, Rapid City, SD 57701, USA}
\affiliation{Dept.~of Physics, University of Wisconsin, River Falls, WI 54022, USA}
\affiliation{Dept.~of Physics and Astronomy, University of Rochester, Rochester, NY 14627, USA}
\affiliation{Oskar Klein Centre and Dept.~of Physics, Stockholm University, SE-10691 Stockholm, Sweden}
\affiliation{Dept.~of Physics and Astronomy, Stony Brook University, Stony Brook, NY 11794-3800, USA}
\affiliation{Dept.~of Physics, Sungkyunkwan University, Suwon 440-746, Korea}
\affiliation{Dept.~of Physics and Astronomy, University of Alabama, Tuscaloosa, AL 35487, USA}
\affiliation{Dept.~of Astronomy and Astrophysics, Pennsylvania State University, University Park, PA 16802, USA}
\affiliation{Dept.~of Physics, Pennsylvania State University, University Park, PA 16802, USA}
\affiliation{Dept.~of Physics and Astronomy, Uppsala University, Box 516, S-75120 Uppsala, Sweden}
\affiliation{Dept.~of Physics, University of Wuppertal, D-42119 Wuppertal, Germany}
\affiliation{DESY, D-15738 Zeuthen, Germany}

\author{M.~G.~Aartsen}
\affiliation{Department of Physics, University of Adelaide, Adelaide, 5005, Australia}
\author{M.~Ackermann}
\affiliation{DESY, D-15738 Zeuthen, Germany}
\author{J.~Adams}
\affiliation{Dept.~of Physics and Astronomy, University of Canterbury, Private Bag 4800, Christchurch, New Zealand}
\author{J.~A.~Aguilar}
\affiliation{Universit\'e Libre de Bruxelles, Science Faculty CP230, B-1050 Brussels, Belgium}
\author{M.~Ahlers}
\affiliation{Niels Bohr Institute, University of Copenhagen, DK-2100 Copenhagen, Denmark}
\author{M.~Ahrens}
\affiliation{Oskar Klein Centre and Dept.~of Physics, Stockholm University, SE-10691 Stockholm, Sweden}
\author{I.~Al~Samarai}
\affiliation{D\'epartement de physique nucl\'eaire et corpusculaire, Universit\'e de Gen\`eve, CH-1211 Gen\`eve, Switzerland}
\author{D.~Altmann}
\affiliation{Erlangen Centre for Astroparticle Physics, Friedrich-Alexander-Universit\"at Erlangen-N\"urnberg, D-91058 Erlangen, Germany}
\author{K.~Andeen}
\affiliation{Department of Physics, Marquette University, Milwaukee, WI, 53201, USA}
\author{T.~Anderson}
\affiliation{Dept.~of Physics, Pennsylvania State University, University Park, PA 16802, USA}
% ...

\date{\today}

\collaboration{IceCube Collaboration}
 \email{analysis@icecube.wisc.edu}
\noaffiliation

\date{\today}

\begin{abstract}
We present a measurement of the atmospheric neutrino oscillation parameters using three years of data from the IceCube Neutrino Observatory.
The DeepCore infill array in the center of IceCube enables detection and reconstruction of neutrinos produced by the interaction of cosmic rays in the Earth's atmosphere at energies as low as $\sim5$~GeV.
That energy threshold permits measurements of muon neutrino disappearance, over a range of baselines up to the diameter of the Earth,
probing the same range of $L/E_\nu$ as long-baseline experiments but with substantially higher energy neutrinos.
This analysis uses neutrinos from the full sky with reconstructed energies from $5.6$~--~$56$~GeV.
We measure $\Delta m^2_{32}=2.31^{+0.11}_{-0.13} \times 10^{-3}$~eV$^2$ and $\sin^2 \theta_{23}=0.51^{+0.07}_{-0.09}$, assuming normal neutrino mass ordering.
These results are consistent with, and of similar precision to, those from accelerator and reactor-based experiments.
\end{abstract}

\pacs{Valid PACS appear here}
\maketitle


\section{\label{sec:introduction}Introduction}

It is well established that the neutrino mass eigenstates do not correspond to the neutrino
flavor eigenstates, leading to flavor oscillations as neutrinos propagate through space~\cite{Fukuda:1998mi,Ahmad:2001an}.
After traveling a distance $L$ a neutrino of
energy $E$ may be detected with a different flavor than it was produced with.
In particular, the muon neutrino survival probability is described approximately by
\begin{equation}
    P(\nu_\mu \rightarrow \nu_\mu) \approx 1 - 4|U_{\mu3}|^2\left(1-|U_{\mu3}|^2\right)\sin^2\left(\frac{\Delta m_{32}^2 L}{4E}\right),
    \label{eq:numu_oscillation}
\end{equation}
where $U_{\mu3}=\sin\theta_{23}\cos\theta_{13}$ is one element of the PMNS~\cite{Pontecorvo:1957cp,Maki:1962mu} matrix $U$ expressed in terms of the mixing angles $\theta_{23}$ and $\theta_{13}$,
$\Delta m^2_{32}=m^2_3-m^2_2$ is the splitting of the second and third neutrino mass states that drives oscillation on the length and energy scales relevant to this analysis.
In addition to the parameters shown in Eq.~\eqref{eq:numu_oscillation}, neutrino oscillations also depend on the parameters $\theta_{12}$, $\Delta m^2_{21}$ and $\delta_{CP}$, but these have a negligible effect on the data presented in this paper.

Interactions of cosmic rays in the atmosphere~\cite{Volkova:1980sw,Barr:2004br,Honda:2015fha}
provide a large flux of neutrinos traveling distances ranging from $L \sim 20$~km (vertically down-going) to
$L \sim 1.3 \times 10^4$~km (vertically up-going) to a detector near the Earth's surface.
For  up-going neutrinos, there is complete muon neutrino disappearance at energies as high as $\sim25$~GeV. Given the density of material traversed by these neutrinos, matter effects alter Eq.~\eqref{eq:numu_oscillation} slightly and must be taken into account~\cite{Wolfenstein:1977ue,Petcov:1998su,Akhmedov:2006hb,Akhmedov:2008qt}.

In this letter, we report our measurement of $\theta_{23}$ and $\Delta m_{32}^2$, using the IceCube Observatory to observe oscillation-induced patterns in the atmospheric
neutrino flux coming from all directions between $5.6$~GeV and $56$~GeV.
The results presented here complement other leading
experiments~\cite{Adamson:2013whj,Abe:2017uxa,Adamson:2017qqn,An:2016ses,Wendell:2014dka} in two ways. 
Long-baseline experiments with baselines of a few hundred kilometers and Super-Kamiokande observe much lower energy events (primarily charged-current quasi-elastic and resonant scattering), while our measurement relies on higher energy deep inelastic scattering events and is thus subject to different sources of systematic uncertainty~\cite{Formaggio:2013kya}.
In addition, the higher energy range of IceCube neutrinos provides complementary constraints on potential new physics in the neutrino sector~\cite{Friedland:2004ah,Friedland:2005vy,Ohlsson:2013epa,Esmaili:2013fva,Gonzalez-Garcia:2013usa,Mocioiu:2014gua,Choubey:2014iia,Coloma:2016gei,Liao:2016hsa,Aartsen:2017bap}.

The IceCube detector was fully commissioned in 2011 and we previously reported results~\cite{Aartsen:2014yll} using data from May 2011 through April 2014.
Those results were obtained using reconstruction tools that relied on unscattered Cherenkov photons and therefore were less susceptible to detector noise.
The results presented here use a new reconstruction that includes scattered photons and retains an order of magnitude more events per year.
Because the detector's noise rates were still stabilizing during the first year of operation,
and the new reconstruction is more susceptible to noise,
we chose before unblinding to use data from April 2012 through May 2015.

\section{\label{sec:detector}The IceCube DeepCore Detector}

The IceCube In-Ice Array~\cite{Aartsen:2016nxy} is composed of 5160 
downward-looking 10" photomultiplier
tubes (PMTs) embedded in a 1~km$^3$ volume of the South Pole glacial ice at depths between 1.45 and 2.45~km.
The PMTs and associated electronics are enclosed in glass pressure
spheres to form digital optical modules (DOMs)~\cite{Abbasi:2008aa,Abbasi:2010vc}.
The DOMs are deployed on 86 vertical strings of 60 modules each.
Of these strings, 78 are deployed in a triangular grid with horizontal spacing of about 125~m between strings.
These DOMs are used primarily as an active veto to reject atmospheric muon
events in this analysis.
The remaining 8 strings fill a more densely instrumented $\sim10^7$~m$^3$ volume of ice in the bottom center of the detector, called DeepCore, enabling detection of neutrinos with energies down to $\sim$5~GeV~\cite{Collaboration:2011ym}.

Neutrino interactions in DeepCore are simulated with GENIE~\cite{Andreopoulos:2009rq}.  Hadrons produced in these interactions are simulated using 
GEANT4~\cite{Agostinelli:2002hh}, as are electromagnetic showers below 100~MeV.  At higher energies, shower-to-shower variation is small enough to permit use of standardized light emission templates~\cite{Radel:2012kw} based on GEANT4 simulations to reduce computation time.
Muons energy losses in the ice are simulated using the PROPOSAL package~\cite{Koehne:2013gpa}.
Cherenkov photons produced by showers and muons are tracked individually using GPU-based software to simulate scattering and absorption~\cite{CLSIM}.

\section{\label{sec:event_selection}Reconstruction and Event selection}

The event reconstruction used in this analysis models the scattering of Cherenkov photons in the ice surrounding our DOMs~\cite{Aartsen:2013rt}
to calculate the likelihood of the observed photoelectrons as a function of the neutrino interaction position, direction, and energy.
Given the complexity of this likelihood space, the MultiNest algorithm~\cite{Feroz:2008xx} is used to find the global maximum.
This reconstruction is run under two different event hypotheses: first a $\nu_\mu$ charged-current (CC) interaction comprising a hadronic shower and collinear muon track emerging from the interaction vertex, and then with only a shower at the vertex (i.e., a nested hypothesis with zero muon track length).  The latter model incorporates $\nu_e$ and most $\nu_\tau$ CC interactions as well as neutral current interactions, as we do not attempt to separate electromagnetic showers produced by a leading lepton from hadronic showers produced by the disrupted nucleus.  

\begin{figure}
    \includegraphics[width=\linewidth]{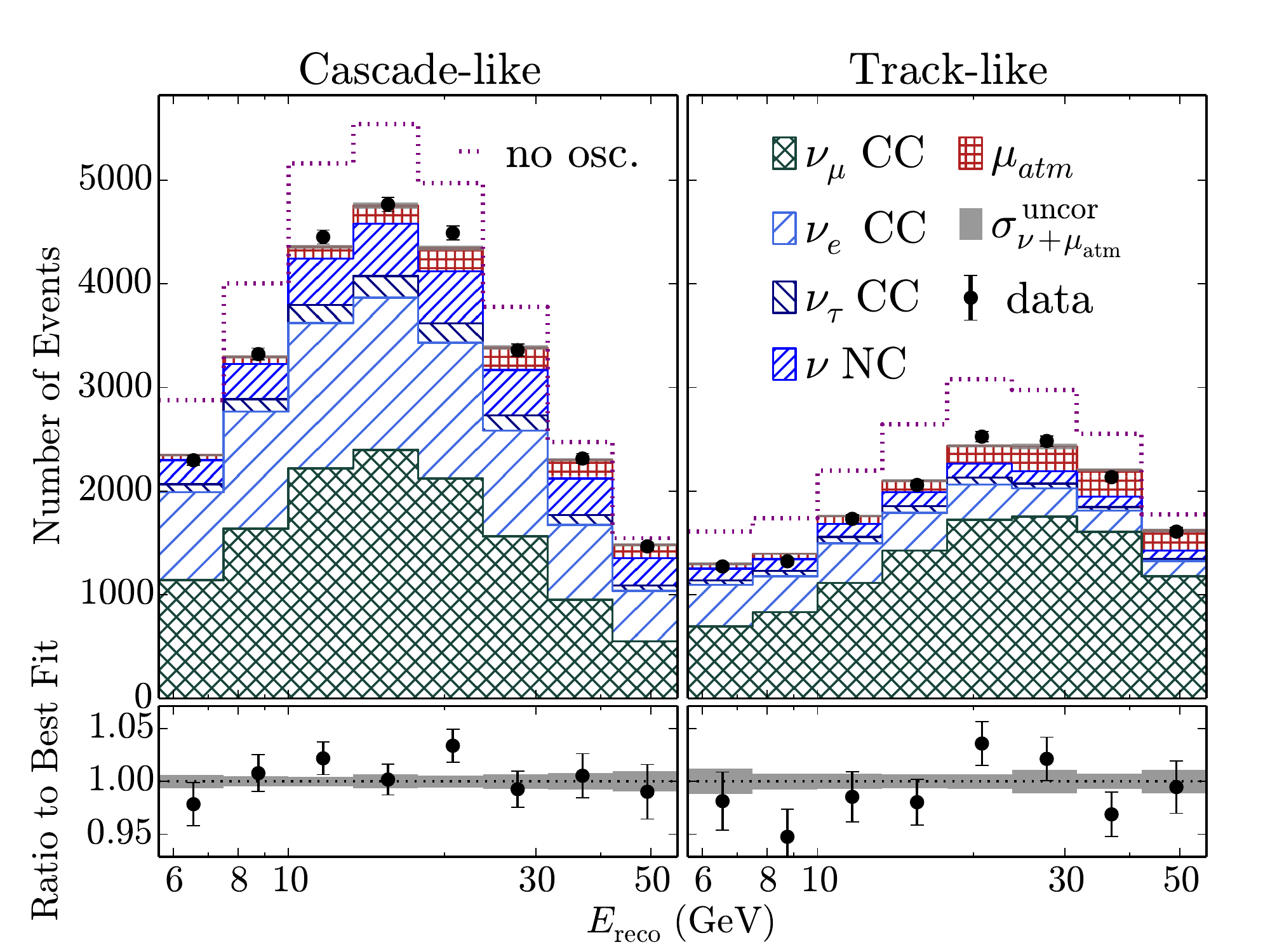}
    \caption{
        Reconstructed energy distributions observed in data (points) and predicted by interaction type at
        our best fit point for oscillations (stacked). In addition to each separate component,
        the uncorrelated statistical uncertainty associated to the expectation
        $\left(\sigma^{\text{uncor}}_{\nu+\mu_{\text{atm.}}}\right)$ is shown in a shaded band.  The track-like sample is peaked at higher energy due to the rising probability of tagging $\nu_\mu$ CC events.  
        The bottom plots show the ratio of the data to the fitted prediction.
    }
    \label{fig:E_plot}
\end{figure}

The $\nu_\mu$~CC reconstruction is used to estimate the direction and energy of the neutrino.  The difference in best-fit
likelihoods between the two hypotheses is used to classify our events as ``track-like,'' if inclusion of a muon track improves the fit substantially, or
``cascade-like,'' if the event is equally well fit without a muon.
The reconstructed neutrino energy ($E_{\text{reco}}$) distributions of events in each of these categories after final selection are shown in Fig.~\ref{fig:E_plot}, along with the corresponding predicted distributions broken down by event type.  
The track-like sample is enriched in $\nu_\mu$~CC events (68\% of sample), especially at higher energies where muons are more likely detected, while the
cascade-like sample is evenly divided between $\nu_\mu$~CC and interactions without
a muon in the final state.
The angular and energy resolutions provided by the reconstruction are energy-dependent, with median resolutions of  10$^\circ$ (16$^\circ$) in zenith angle and 24\% (29\%) in neutrino energy for track-like (cascade-like) events at $E_\nu = 20$~GeV.

The event selection in this analysis uses the DOMs
surrounding the DeepCore region to veto atmospheric muons.
The first criteria remove accidental triggers caused by dark noise by demanding a minimum amount of light detected in the DeepCore volume, with timing and spatial scale consistent with a particle emitting Cherenkov radiation.  Events in which photons are observed outside the DeepCore volume before the light detected inside DeepCore, in a time window consistent with atmospheric muons penetrating to the fiducial volume, are then rejected. 
These are followed by a boosted decision tree (BDT)~\cite{Hocker:2007ht} which further reduces the background of atmospheric muons.  The BDT uses the timing and spatial scale of the detected photoelectrons to select events with substantial charge deposition at the beginning of the event, indicative of a neutrino interaction vertex.  It also considers how close the event is to the border of the DeepCore volume and the results of several fast directional reconstructions~\cite{Ahrens:2003fg} in determining whether the event may be an atmospheric muon.
Finally, we demand that the interaction vertex reconstructed by the likelihood fit described above be contained within DeepCore and the end of the reconstructed muon be within the first row of DOMs outside DeepCore, which further reduces atmospheric muon contamination and improves reconstruction accuracy.

As these selection criteria reduce the atmospheric muon rate by a factor of approximately $10^{8}$, it is challenging to simulate enough atmospheric muons to obtain a reliable prediction for the distribution of the remaining muons, especially in the presence of systematic uncertainties.  We instead use a data-driven 
estimate of the shape of the muon background distributions, with the normalization free to float.  This approach is based on tagging events that would have been accepted except for a small number of photons detected in the veto region, similar to the procedure in 
Ref.~\cite{Aartsen:2014yll}.  The uncertainty in the background shape is estimated using two different criteria for tagging these events, and was compared to the currently available muon Monte Carlo. This uncertainty is added in quadrature to the statistical
uncertainties in the tagged background event sample and the neutrino Monte Carlo, to provide the total uncorrelated statistical uncertainty 
$\left(\sigma^{\text{uncor}}_{\nu+\mu_{\text{atm.}}}\right)$ in the expected distribution shown in Fig.~\ref{fig:E_plot}.

\section{\label{sec:analysis}Analysis}

The final fit of the data is done using an $8\times8\times2$ binned histogram, with 8 bins in $\log_{10}{{E}_\text{reco}}$, 8 bins in the cosine of the reconstructed neutrino zenith direction ($\cos{\theta_{z,\text{reco}}}$), one track-like  bin and one cascade-like.
The bins are equally spaced with $\cos{\theta_{z,\text{reco}}} \in [-1,1]$ and $\log_{10}{E_\text{reco}} \in [0.75, 1.75]$.
The fit assumes three-flavor oscillations with $\Delta m^2_{21}=7.53\times 10^{-5}$~eV$^2$, $\sin^2 \theta_{12}=0.304$, $\sin^2 \theta_{13}=2.17\times 10^{-2}$, and $\delta_{CP}=0^{\circ}$.  

We use MINUIT2~\cite{James:1975dr} to minimize a function
\begin{equation}
    \chi^2 = \sum_{i \in \{\text{bins}\}} \frac{(n^{\nu+\mu_{\text{atm}}}_i - n^{\text{data}}_i)^2}{(\sigma^{\text{data}}_i)^2 + (\sigma^{\text{uncor}}_{\nu+\mu_{\text{atm}},i})^2} + \sum_{j \in \{\text{syst}\}} \frac{(s_j - \hat{s}_j)^2}{\hat{\sigma}_{s_j}^2},
\label{eq:chi2}
\end{equation}
where $n^{\nu+\mu_{\text{atm}}}_i$ is the number of events expected in the $i^{th}$ bin,
which is the sum of neutrino events weighted to the desired oscillation parameters using Prob3++~\cite{prob3pp} 
and the atmospheric muon background.
The number of events observed in the $i^{th}$ bin is $n^{\text{data}}_i$, with Poisson uncertainty
$\sigma^{\text{data}}_i=\sqrt{n^{\text{data}}_i}$, and
$\sigma^{\text{uncor}}_{\nu+\mu_{\text{atm}},i}$ is the uncertainty in the prediction of the number of events of the $i^{th}$ bin.
$\sigma^{\text{uncor}}_{\nu+\mu_{\text{atm}}}$ includes both effects of finite MC statistics and uncertainties in our data-driven muon background estimate.
The second term of Eq.~\eqref{eq:chi2} is a penalty term for our nuisance parameters, where $s_j$ is the value of $j^{th}$ systematic, $\hat{s}_j$ is the central value and $\hat{\sigma}_{s_j}^2$ is the Gaussian width of the $j^{th}$ systematic prior.

The analysis includes eleven nuisance parameters describing our systematic uncertainties, summarized in Table~\ref{tab:syst}.  
Seven of these are related to systematic uncertainties in the atmospheric neutrino flux and interaction cross sections. Since only the event rate is observed directly, some uncertainties in flux and cross section have similar effects on the data.  In these cases, the degenerate effects are combined into a single parameter.  Because analytical models of these effects are available, these parameters can be varied continuously by reweighting simulated events. 

The first nuisance parameter is the overall normalization of the event rate. It is affected by uncertainties in the atmospheric neutrino flux and the neutrino interaction cross section, and by the possibility of accidentally vetoing neutrino events due to unrelated atmospheric muons detected in the veto volume.  This last effect is expected to reduce the neutrino rate by several percent, but is not included in the present simulations.
Because of this and the fact it encompasses several effects, no prior is used for this parameter.

A second parameter allows an energy-dependent shift in the event rate.  This can arise from uncertainties in either the spectral index of the atmospheric flux (nominally $\gamma=-2.66$ at the relevant energies in our neutrino flux model~\cite{Honda:2015fha}), or the deep inelastic scattering (DIS) cross section.  A prior of $\hat{\sigma}_{s}=0.10$ is placed on the spectral index to describe the range of these uncertainties.  

Several uncertainties on the DIS cross section were implemented in the fit, but found  either to have negligible impact or to be highly degenerate with the normalization and spectral index parameters over the energy range of this analysis.  
These include variation of parameters of the Bodek-Yang model~\cite{Bodek:2002ps} used in GENIE, uncertainties in the differential DIS cross-section, and hadronization uncertainties for high-$W$ DIS events ~\cite{Katori:2016blu}.  As these effects are captured by the first two nuisance parameters, the additional parameters were not used.  

One neutrino cross-section uncertainty was not well described by these parameters: the uncertainty of the axial mass form factor for resonant events.
The default value of 1.12~GeV and prior of 0.22~GeV were taken from GENIE~\cite{Andreopoulos:2009rq}.  
Uncertainties in CCQE interactions were also investigated but had no impact on the analysis due to the small percentage of CCQE events at these energies.

The normalizations of $\nu_{e} + \bar{\nu}_e$ events and NC events, defined relative to $\nu_{\mu} + \bar{\nu}_{\mu}$ CC events, are both assigned an uncertainty of 20\%.
Uncertainties in hadron production (especially pions and kaons) in air showers affect the predicted flux, in particular the ratio of neutrinos to anti-neutrinos.  We model these hadronic flux effects with two parameters, one dependent on neutrino energy and the other on the zenith angle, chosen to reproduce the uncertainties estimated in Ref.~\cite{Barr:2006it}. 
Their total uncertainty varies from 3\%-10\% depending on the energy and zenith angle, so the fit result is given in units of $\sigma$ as calculated by Barr et al.  Uncertainties in the relative cross section of neutrinos vs.~anti-neutrinos are degenerate with the flux uncertainty in this energy range.
 
\begin{table}
    \caption{\label{tab:syst} Table of nuisance parameters along with their associated priors, if applicable.  The right two columns show the results from our best fit for normal mass ordering and inverted mass ordering, respectively.}
    \begin{tabular}{lccc}
        \hline \hline
        \multirow{2}{*}{Parameters} & \multirow{2}{*}{Priors} & \multicolumn{2}{c}{Best Fit} \\
         & & NO & IO \\
        \hline \multicolumn{4}{c}{Flux and cross section parameters} \\ \hline
		Neutrino event rate [\% of nominal] & no prior & 85 & 85 \\
		$\Delta \gamma$ (spectral index) & 0.00$\pm$0.10 & -0.02           & -0.02 \\
        $M_A$ (resonance) [GeV] & 1.12$\pm$0.22 & 0.92 & 0.93 \\
        $\nu_e + \bar{\nu}_e$ relative normalization [\%] & 100$\pm$20 & 125 & 125 \\
        NC relative normalization [\%] & 100$\pm$20 & 106 & 106 \\
        Hadronic flux, energy dependent [$\sigma$] & 0.00$\pm$1.00 & -0.56 & -0.59 \\
        Hadronic flux, zenith dependent [$\sigma$] & 0.00$\pm$1.00 & -0.55 & -0.57 \\
        \hline \multicolumn{4}{c}{Detector parameters} \\ \hline
        overall optical eff. [\%] & 100$\pm$10 & 102 & 102 \\
        relative optical eff., lateral [$\sigma$] & 0.0$\pm$1.0 & 0.2 & 0.2 \\
        relative optical eff., head-on [a.u.] & no prior & -0.72 & -0.66 \\
        \hline \multicolumn{4}{c}{Background} \\ \hline
        Atm. $\mu$ contamination [\% of sample] & no prior & 5.5 & 5.6 \\
        \hline \hline
    \end{tabular}
\end{table}

Systematics related to the response of the detector itself, including photon propagation through the ice and the anisotropic sensitivity of the DOMs, have the largest impact on this analysis.  
Their effects are estimated by Monte Carlo simulation at discrete values,
with the contents of each
bin in the (energy, direction, track/cascade) analysis histogram determined by linear interpolation between the discrete simulated models, following the approach of Ref.~\cite{Aartsen:2014yll,Aartsen:2017bap}.

Uncertainties in the efficiency of photon detection 
are driven by the formation of bubbles in the refrozen ice columns in the holes where the IceCube strings were deployed. 
A prior with a width of 10\% was applied to the overall photon collection efficiency~\cite{Aartsen:2016nxy}, parametrized using seven MC data sets ranging from 88\% to 112\% of the nominal optical efficiency.
In addition to modifying the absolute efficiency, these bubbles can scatter Cherenkov photons near the DOMs, modulating the relative optical efficiency as function of the incident photon angle.  
The effect of the refrozen ice column is modeled by two effective parameters controlling the shape of the DOM angular acceptance curve.

The first parameter controls the lateral angular acceptance (i.e., relative sensitivity to photons traveling roughly 20$^\circ$ above versus below the horizontal) and is fairly well constrained by LED calibration data.
Five MC data sets were generated covering the $-1\sigma$ to $+1\sigma$ uncertainty from the LED calibration, and parametrized in the same way as the overall optical efficiency described above.
A Gaussian prior based on the LED data is used.

The second parameter controls sensitivity to photons traveling vertically upward and striking the DOMs head-on, which is not well constrained by string-to-string LED calibration.
That effect is modeled using a dimensionless parameter ranging from -5 (corresponding to a bubble column completely obscuring the DOM face for vertically incident photons) to $2.5$ (no obscuration).  Zero corresponds to constant sensitivity for angles of incidence from $0^\circ$ to $30^\circ$ from vertical.
Six MC sets covering the range from -5 to 2 were used to parametrize this effect.
No prior is applied to this parameter due to lack of information from calibration data.

The last nuisance parameter controls the level of atmospheric muon contamination in the final sample.
As described above, the shape of this background in the analysis histogram, including binwise uncertainties, is derived from data.  Since the absolute efficiency for tagging background events with this method is unknown, the normalization of the muon contribution is left free in the fit.

In addition to the systematic uncertainties discussed above,
we have considered the impact of seed dependence in our event reconstruction,
different optical models for both the undisturbed ice and the refrozen ice columns, and
an improved detector calibration currently being prepared.
In all these cases the impact on the final result was found to be minor,
and they were thus omitted from the fit and the error estimate.

\section{\label{sec:results}Results and Conclusion}

The 
analysis procedure described above gives a best fit of $\Delta m^2_{32}=2.31^{+0.11}_{-0.13} \times 10^{-3}$~eV$^2$ and $\sin^2 \theta_{23}=0.51^{+0.07}_{-0.09}$, assuming normal neutrino mass ordering (NO).
For the inverted mass ordering (IO), the best fit shifts to $\Delta m^2_{32}=-2.32 \times 10^{-3}$~eV$^2$ and $\sin^2 \theta_{23}=0.51$.
The pulls on the nuisance parameters are shown in Table~\ref{tab:syst}.
Though IceCube's current sensitivity to the mass ordering is low, dedicated analyses are underway to measure this.

The data agree well with the best-fit MC data set, with $\chi^2 = 117.4$ for both neutrino mass orderings.
This corresponds to a $p$-value of 0.52 given the 119 effective degrees of freedom estimated
via toy MCs, following the procedure described in Ref.~\cite{Aartsen:2017bap}.

\begin{figure}
    \includegraphics[width=\linewidth]{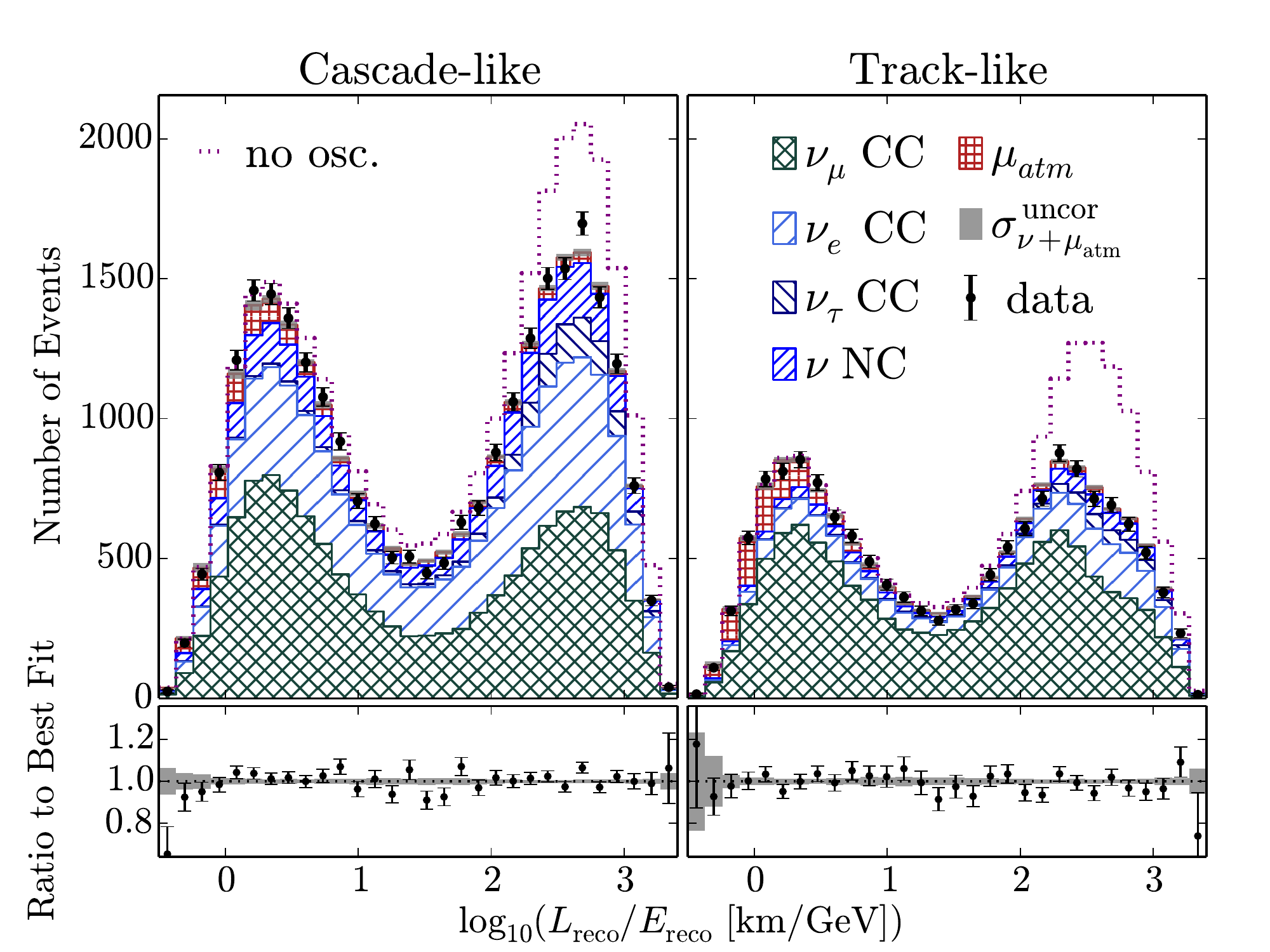}
    \caption{Data projected onto $L/E$ for illustration.
    The black dots indicate the data along with their corresponding statistical errors.
    The dotted line shows the expectation in the absence of neutrino oscillations.
    The stacked hatched histograms are the predicted counts given the best-fit values of all parameters in the fit for each component.
    The bottom plots show the ratio of the data to the fitted prediction.  The bars indicate statistical uncertainties and shaded region the $\sigma^{\text{uncor}}_{\nu+\mu_{\text{atm}}}$ uncertainty in the expectation, as defined in Eq.~\eqref{eq:chi2}, which is dominated by the uncertainty in $\mu_{atm}$.      }
    \label{fig:LE_plot}
\end{figure}

To better visualize the fit, Fig.~\ref{fig:LE_plot} shows the results of the fit projected onto a single $L/E$ axis, for both the track-like and cascade-like events.  The two peaks in each distribution correspond to down-going and up-going neutrino trajectories.  Up-going $\nu_\mu + \bar{\nu}_\mu$ are strongly suppressed in the track-like channel due to oscillations.  Some suppression of up-going cascade-like data is also visible, due to disappearance of lower-energy $\nu_\mu$ which are not tagged as track-like by our reconstruction.  

\begin{figure}
    \includegraphics[width=\linewidth]{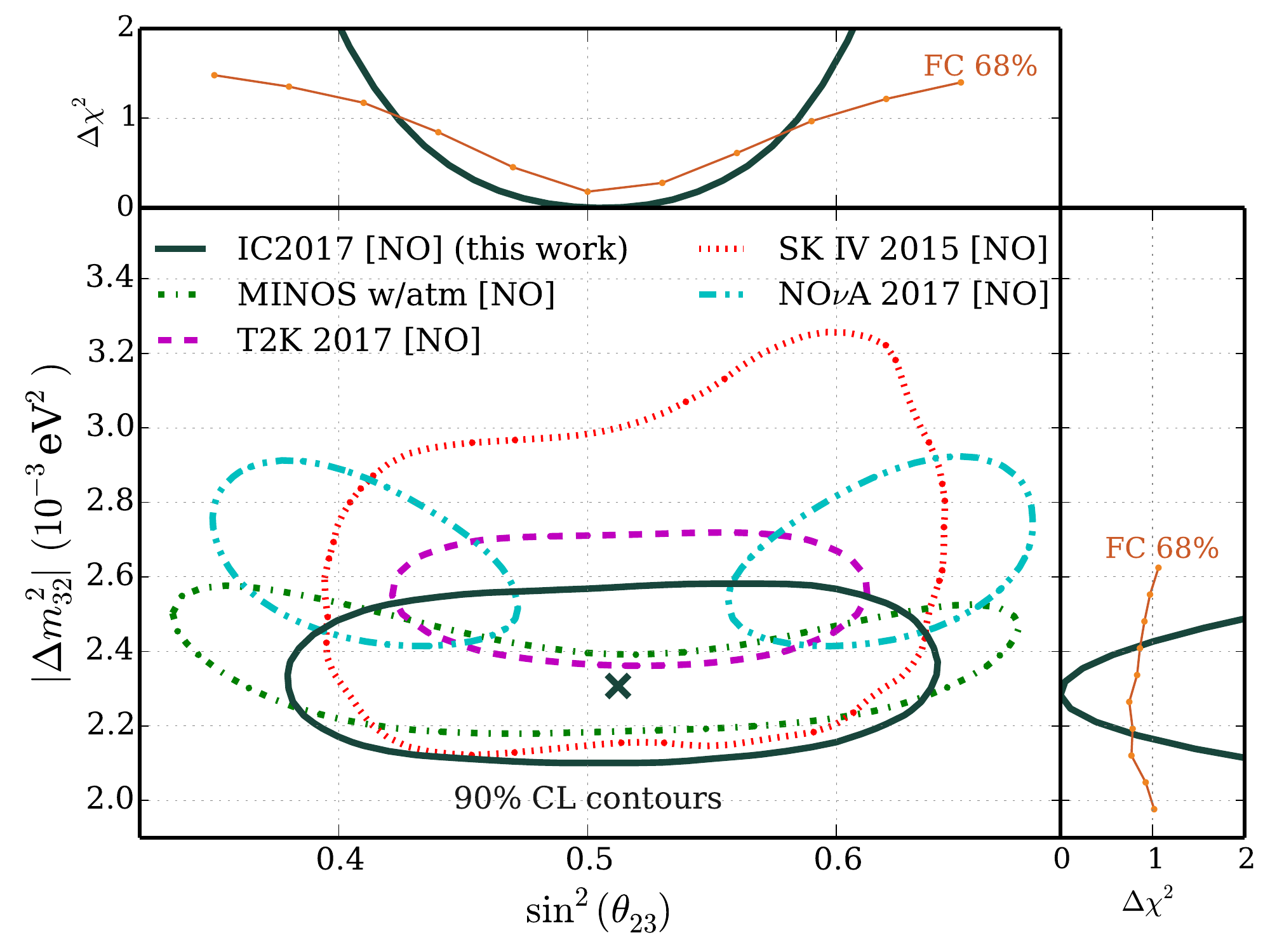}
    \caption{The 90\% allowed region from this work (solid line) compared to other experiments~\cite{Abe:2017uxa,Wendell:2014dka,Adamson:2013whj,Adamson:2017qqn} (dashed lines).  The cross marks our best-fit point.  The outer plots show the results of the 1-D projections after profiling over the other variables along with the 68\% 
CL $\Delta \chi^2_c$ threshold estimated using the Feldman-Cousins method~\cite{Feldman:1997qc}.}
    \label{fig:contours}
\end{figure}

Figure~\ref{fig:contours} shows the region of $\sin^2 \theta_{23}$ and $\Delta m^2_{32}$ allowed by our analysis at 90\% C.L., along with our best fit and several other leading measurements of these parameters~\cite{Abe:2017uxa,Wendell:2014dka,Adamson:2013whj,Adamson:2017qqn}.
The contours are calculated using the approach of Feldman and Cousins~\cite{Feldman:1997qc} to ensure proper coverage.

Our results are consistent with those from other
experiments~\cite{Abe:2017uxa,Wendell:2014dka,Adamson:2013whj,Adamson:2017qqn,An:2016ses}, but using significantly higher energy neutrinos and subject to a different set of systematic uncertainties.  
Our data prefer maximal mixing, similar to the result from T2K~\cite{Abe:2017uxa}.  The best-fit values from the NO$\nu$A experiment~\cite{Adamson:2017qqn} are disfavored by $\Delta \chi^2=8.9$ (first octant) or $\Delta \chi^2=8.8$ (second octant), corresponding to a significance of $2.6\sigma$ using the method of Feldman and Cousins, although there is considerable overlap in the 90\% confidence regions of the two measurements.
Further improvements to our analysis are underway, including the incorporation of additional years of data, extensions of our event selections, and improved calibration of the detector response.

\section*{Acknowledgements}

\begin{acknowledgments}

We acknowledge the support from the following agencies:
U.S. National Science Foundation-Office of Polar Programs,
U.S. National Science Foundation-Physics Division,
University of Wisconsin Alumni Research Foundation,
Michigan State University,
the Grid Laboratory Of Wisconsin (GLOW) grid infrastructure at the University of Wisconsin - Madison, the Open Science Grid (OSG) grid infrastructure;
U.S. Department of Energy, and National Energy Research Scientific Computing Center,
the Louisiana Optical Network Initiative (LONI) grid computing resources;
Natural Sciences and Engineering Research Council of Canada,
WestGrid and Compute/Calcul Canada;
Swedish Research Council,
Swedish Polar Research Secretariat,
Swedish National Infrastructure for Computing (SNIC),
and Knut and Alice Wallenberg Foundation, Sweden;
German Ministry for Education and Research (BMBF),
Deutsche Forschungsgemeinschaft (DFG),
Helmholtz Alliance for Astroparticle Physics (HAP),
Initiative and Networking Fund of the Helmholtz Association,
Germany;
Fund for Scientific Research (FNRS-FWO),
FWO Odysseus programme,
Flanders Institute to encourage scientific and technological research in industry (IWT),
Belgian Federal Science Policy Office (Belspo);
Marsden Fund, New Zealand;
Australian Research Council;
Japan Society for Promotion of Science (JSPS);
the Swiss National Science Foundation (SNSF), Switzerland;
National Research Foundation of Korea (NRF);
Villum Fonden, Danish National Research Foundation (DNRF), Denmark

\end{acknowledgments}

\bibliographystyle{apsrev}
\bibliography{bib/osc}

\beginsupplement
\clearpage
\include{supplemental_material}

\end{document}

%% file: supplemental_material.tex
%
%
%
%

%
%




\graphicspath{{./}{./supplemental_figures/}}





\section{Supplemental Material}

Control of systematic uncertainties in this analysis relies
fundamentally on the use of the full 3D space of neutrino energy,
arrival direction (correlated with path length through the Earth), and
particle type to disentangle systematic
uncertainties from the neutrino
oscillation physics of interest.  Oscillation effects have a distinctive shape in the
$L/E_\nu$ space and primarily affect $\nu_\mu$ CC events, while systematic
effects have a much broader impact on the data.  A complete description of our methodology will be
included in a more detailed forthcoming paper which extends this
analysis to measure the rate of $\nu_\tau$ appearance.  In this
supplement, we provide an abbreviated discussion to illustrate the
approach. 

\begin{figure}
  \centering
  \includegraphics[width=0.8\linewidth]{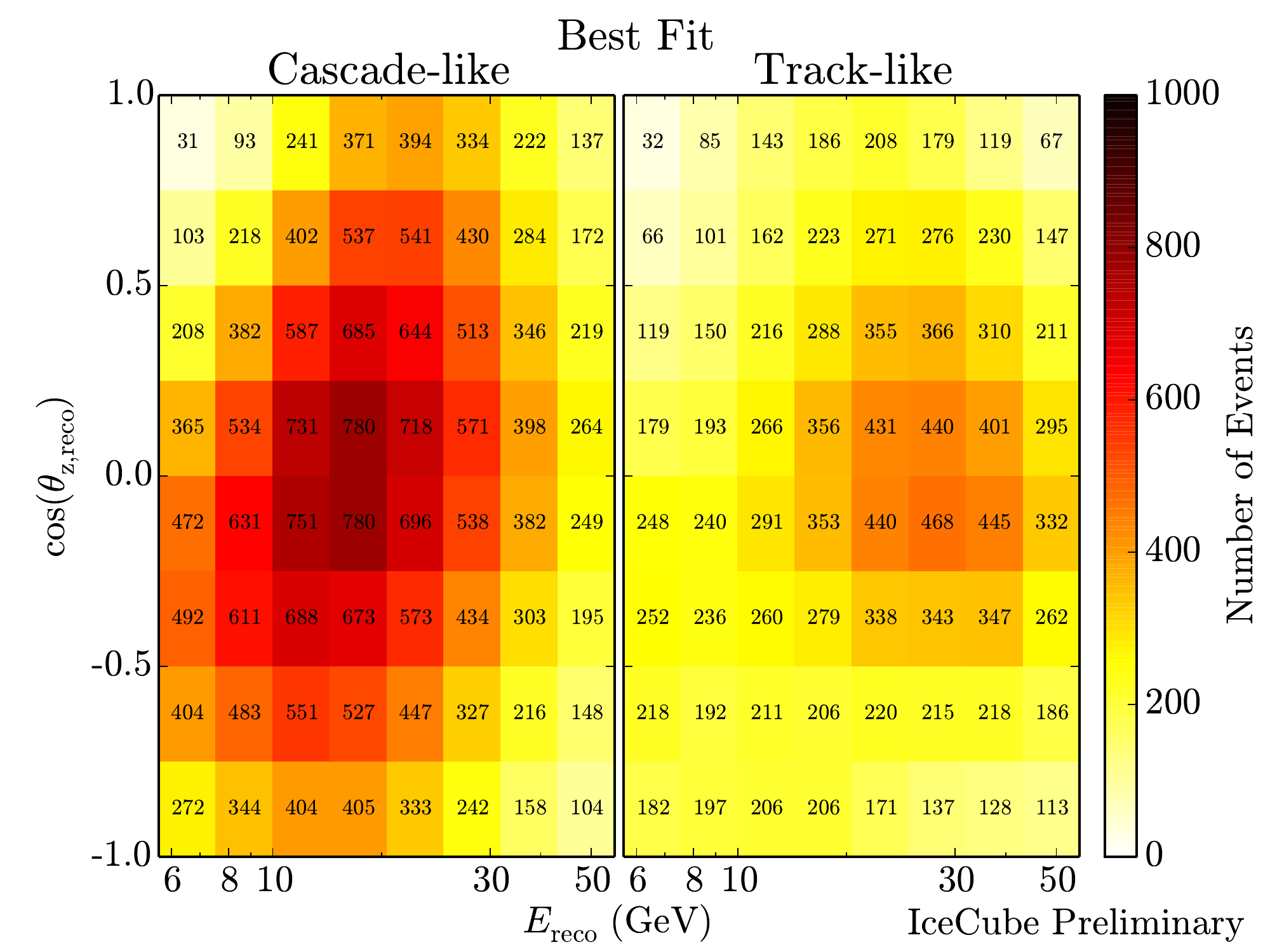}
  \caption{The total number of events observed in each bin of
    reconstructed neutrino energy vs.~zenith angle, which is
    proportional to the path length through the Earth.
    Cascade-like events are shown on the left, track-like events on
    the right.
  }
  \label{fig:bestfit_bin}
\end{figure}

This analysis divides the data into 128 bins in three dimensions: eight
bins in $\log_{10} (E_{\nu,\mathrm{reco}})$, eight bins in
$\cos({\theta_{z,reco}})$, and two bins for  particle identification,
track-like and cascade-like.  Ideally, $\nu_\mu$ CC events are
classified as track-like while $\nu_e$ $\nu_\tau$, and
NC events should be classified as cascade-like; in practice there is
leakage between the samples due to imperfect particle identification.
In our letter, we provide  
projections of the underlying data so that the oscillatory behavior in
reconstructed $L/E_\nu$ and the energy range of the data set may be
seen clearly.  The full 3D distribution of the data is shown in Fig.~\ref{fig:bestfit_bin}.

\begin{figure}
  \centering
  \includegraphics[width=0.8\linewidth]{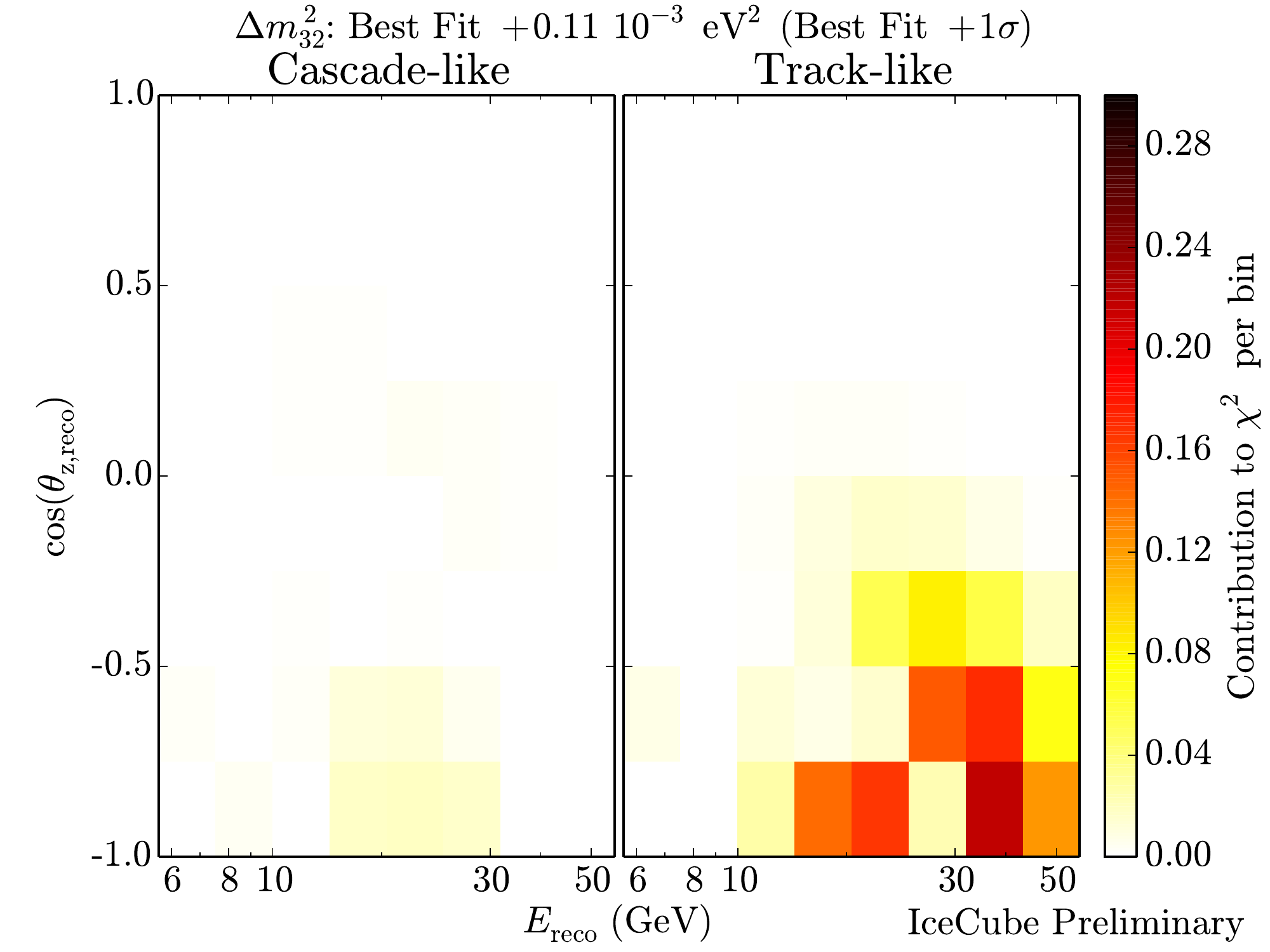}
  \caption{The contribution per analysis bin to the $\chi^2$ of the
    fit if $\Delta m^2_{32}$ were increased by 0.11$\times10^{-3}$
    eV$^2$, corresponding to the $1\sigma$ uncertainty in the fit.
  }
  \label{fig:chi2_signal_dm2}
\end{figure}

\begin{figure}
 \centering 
  \includegraphics[width=0.8\linewidth]{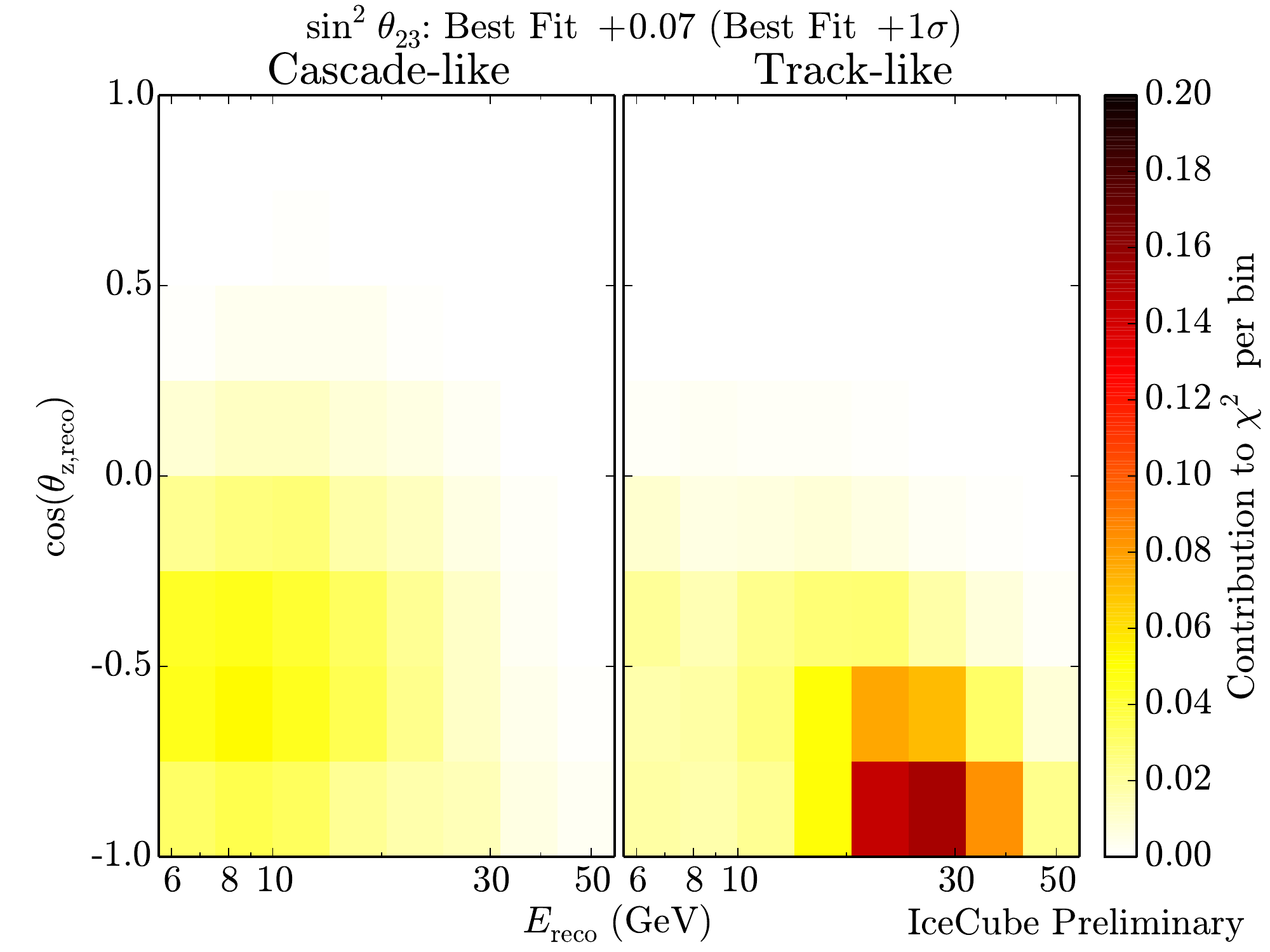} 
  \caption{The contribution per analysis bin to the $\chi^2$ of the
    fit if $\sin^2(\theta_{23})$were increased by 0.07, corresponding
    to the $1\sigma$ uncertainty in the fit.
  }
  \label{fig:chi2_signal_sinth}
\end{figure}

\begin{figure}
  \centering
  \includegraphics[width=0.8\linewidth]{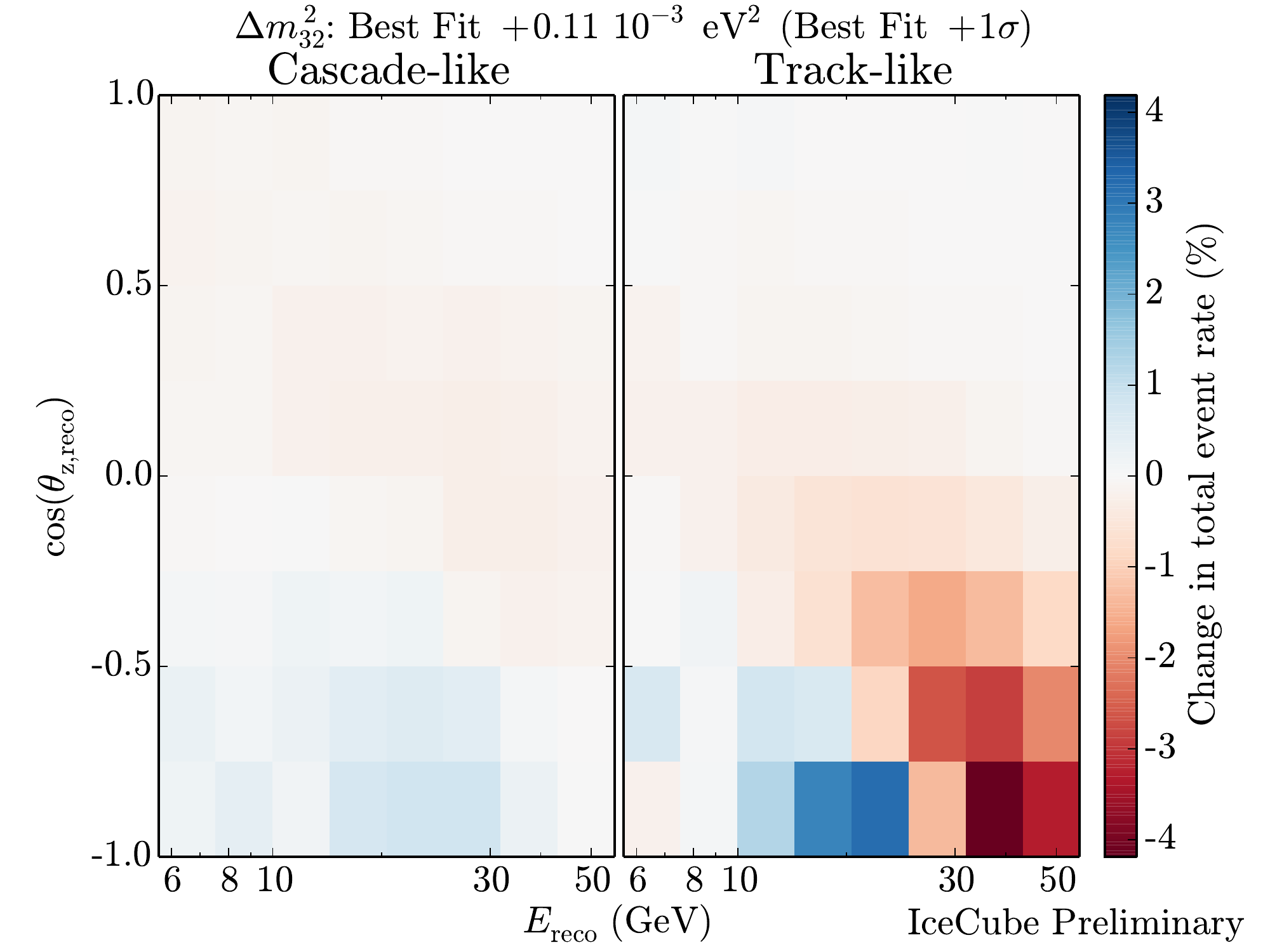}
  \caption{The relative change in event rate per analysis bin if $\Delta m^2_{32}$ were increased by 0.11$\times10^{-3}$
    eV$^2$, corresponding to the $1\sigma$ uncertainty in the fit.
  }
  \label{fig:ratio_dm31}
\end{figure}

\begin{figure}
 \centering 
  \includegraphics[width=0.8\linewidth]{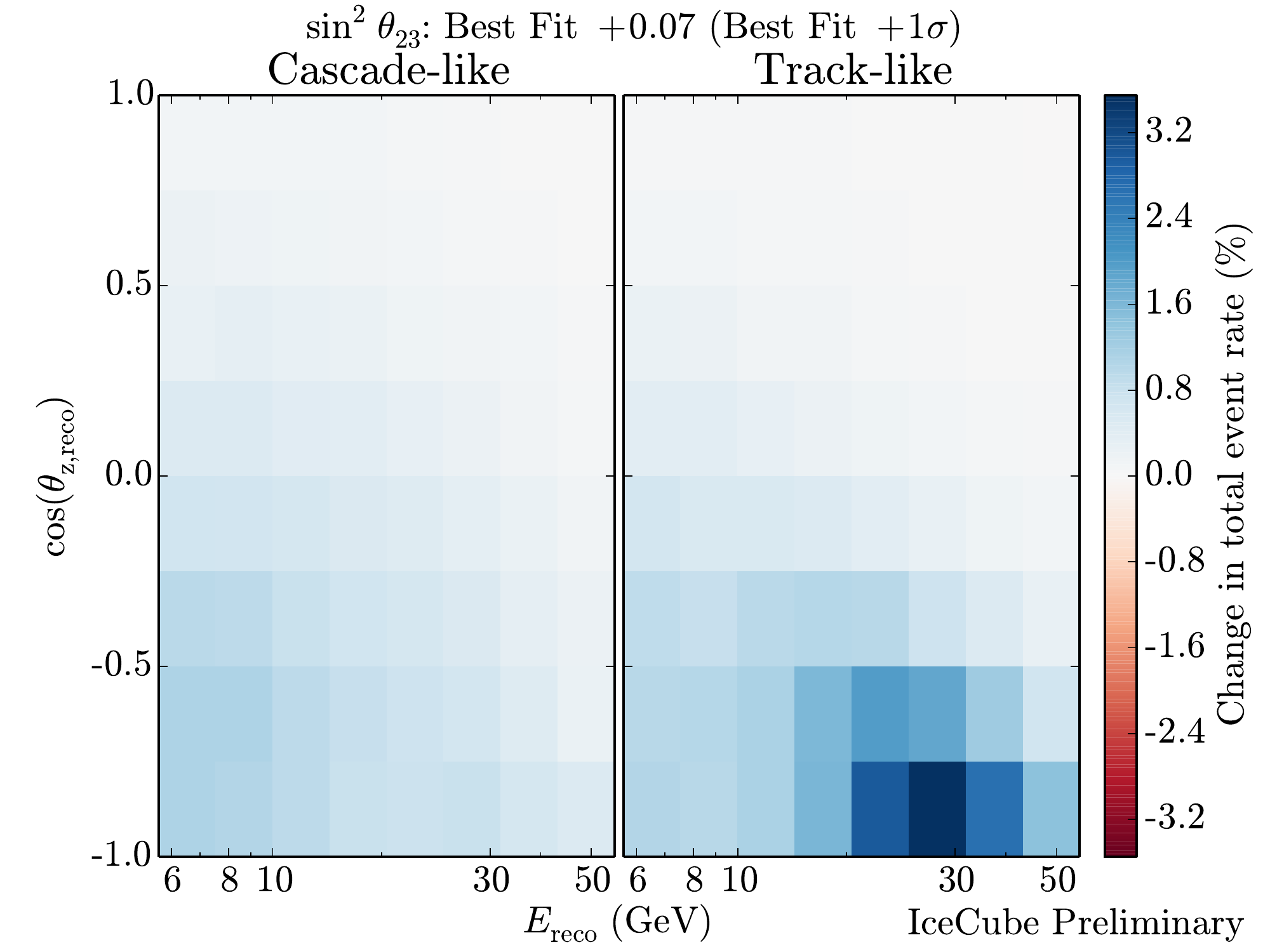} 
  \caption{The relative change in event rate per analysis bin if $\sin^2(\theta_{23})$were increased by 0.07, corresponding
    to the $1\sigma$ uncertainty in the fit.
  }
  \label{fig:ratio_theta}
\end{figure}

Oscillations affect upward-going $\nu_\mu$ CC
events, which are enriched in the track-like sample but also
contribute to the cascade-like sample (especially at lower
reconstructed energy).   
Downward-going neutrinos with baselines too short for oscillations to
occur and cascade-like events provide constraints
on systematic uncertainties related to the atmospheric flux, neutrino
interactions, and detector response.  Our measurement of the
oscillation parameters is obtained mainly from the higher energy bins, as shown in
Figs.~\ref{fig:chi2_signal_dm2} and \ref{fig:chi2_signal_sinth}, due
to the improved angular and energy resolution 
and particle identification accuracy we obtain at higher energies. 
These figures shows how the $\chi^2$ sum would change if the values of $\Delta m^2_{32}$ 
and $\sin^2(\theta_{23})$ were increased by $1\sigma$ from their
best-fit values.   The relative effects on the event rate itself are
shown in Figs.~\ref{fig:ratio_dm31} and \ref{fig:ratio_theta}.  Note
that while a number of $\nu_\mu$ CC events are incorrectly
reconstructed as cascade-like, as shown in Fig.~2 of the letter, their
impact on the measurement of the oscillation parameters is relatively
small compared to the high-energy track-like sample.

The patterns shown in Figs.~\ref{fig:chi2_signal_dm2} and
\ref{fig:chi2_signal_sinth} are the oscillation signature that must be
distinguished from the effects of systematics.
There are eleven different systematic uncertainties implemented in
this analysis, each with their own distinctive signature.  Crucially,
all of these effects have a much broader impact on the data set than
the oscillation physics -- large numbers of bins in both the cascade-like
and track-like samples are affected.
 
\begin{figure}
  \centering
  \includegraphics[width=0.8\linewidth]{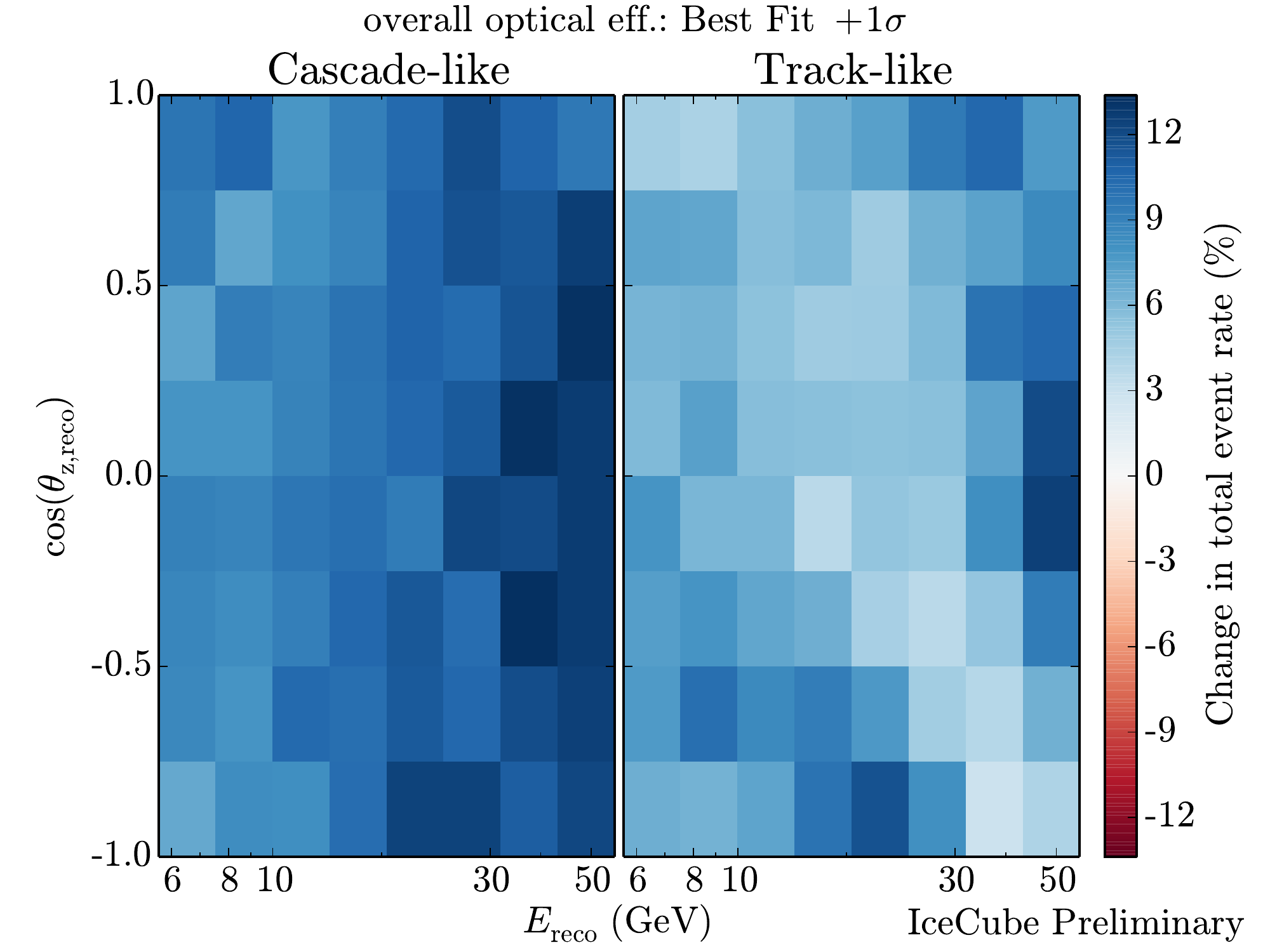} 
    \caption{The relative change in event rate per analysis bin when increasing the overall optical efficiency by 10\%.
    }
    \label{fig:ratio_optical}
\end{figure}

\begin{figure}
  \centering
  \includegraphics[width=0.8\linewidth]{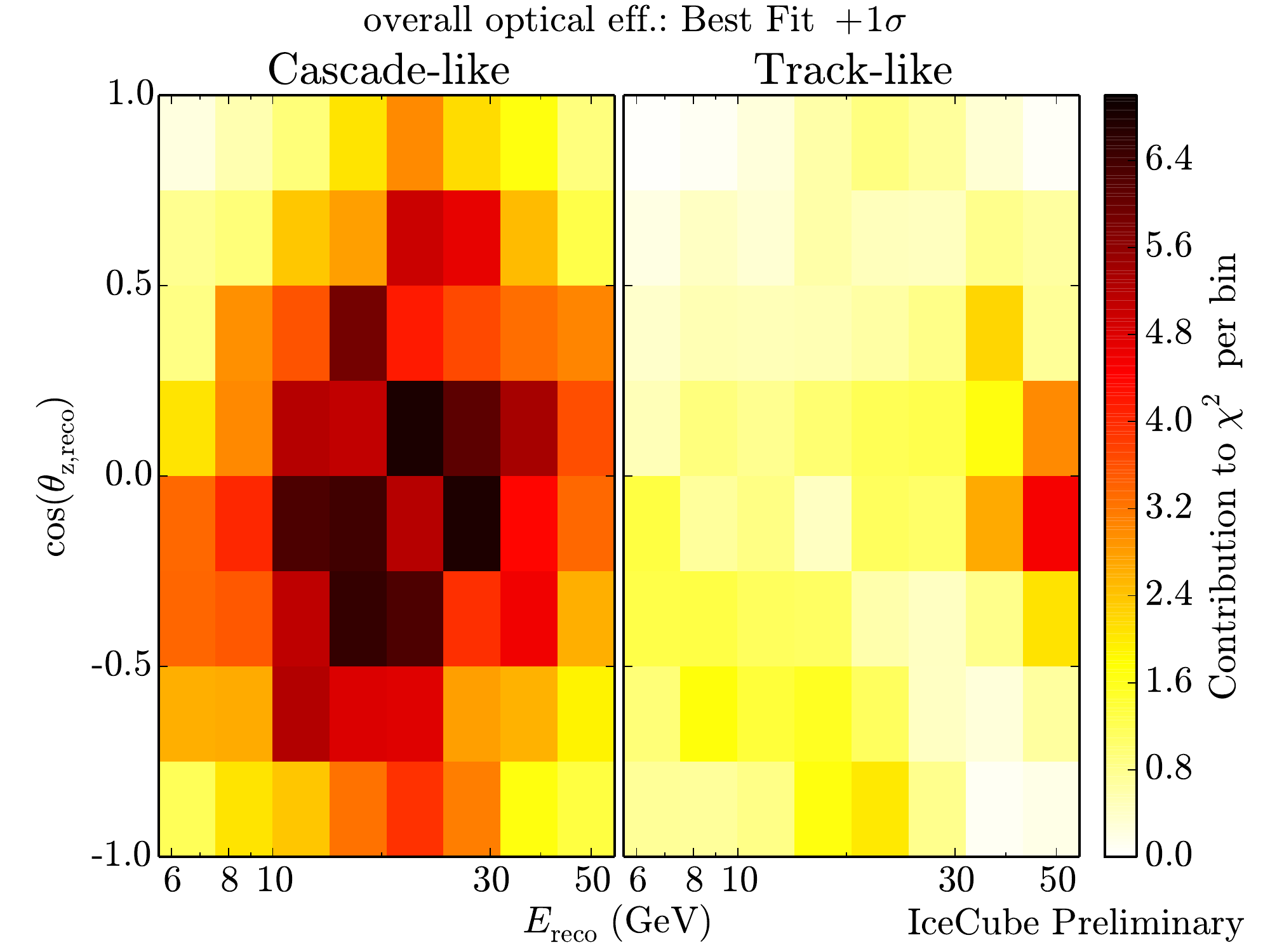} 
    \caption{The contribution per analysis bin to the $\chi^2$ of the
    fit if the overall optical efficiency were increased by 10\%.
    The constraint on this parameter comes almost entirely from the
    cascade-like events, with very little contribution from the bins
    crucial for the measurement of oscillation parameters.
    }
    \label{fig:chi2_optical}
\end{figure}

As an example, the impact of a shift in the overall optical
efficiency of the DOMs, and the range of bins which constrain this
effect, are shown in Figs.~\ref{fig:ratio_optical} and \ref{fig:chi2_optical}.  The
figure shows how a 10\% increase in collection efficiency for
Cherenkov photons (the width of the {\it
  a priori} uncertainty in this parameter) would affect the event
distribution, with all other parameters held fixed.  If only
upward-going track-like events in the 10--40~GeV range were considered, there
would be strong correlation with $\sin^2(\theta_{23})$: higher
optical efficiency produces an increase in the event rate, which partly
fills in the disappearance minimum.   
However, the effect of varying optical efficiency extends throughout both the
cascade-like and track-like distributions, affecting both
upward-going and downward-going events at all energies, while the
effect of oscillations 
is well localized to the 10--40~GeV range at angles of
$\cos(\theta_{z,\textrm{reco}}) < -0.6$, as shown in
Fig.~\ref{fig:ratio_theta}.   Consideration of the full energy
vs.~angle distribution of both samples thus allows the two effects to
be disentangled effectively.  As shown in Fig.~\ref{fig:chi2_optical},
the constraint on this systematic arises almost entirely from the
cascade sample.  Due to the correlations with oscillation parameters,
the fitter does not obtain constraints on this systematic from
up-going high-energy $\nu_\mu$ tracks.

\begin{figure}
  \includegraphics[width=\linewidth]{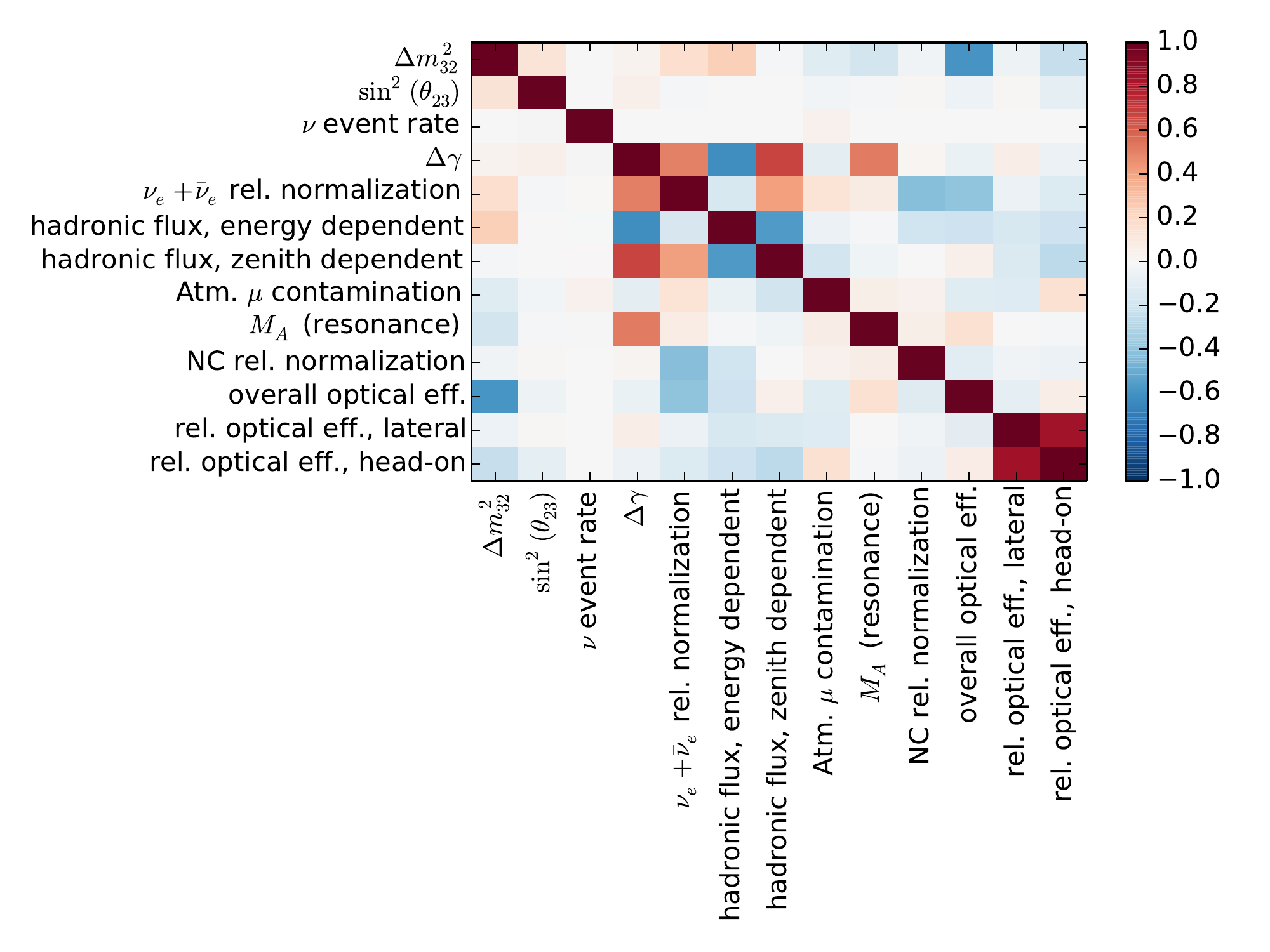}
  \caption{Correlation coefficients from the simultaneous fit of the eleven systematic
    parameters and the two oscillation parameters.}
  \label{fig:correlations}
\end{figure}

Affecting the measurement of $\Delta m^2_{32}$ is particularly
complicated, since it requires shifting the position of the
oscillation dip.  Such a shift requires both
increases and decreases in event rate  
in neighboring bins, while leaving most of the distribution
unaffected, as shown in Fig.~\ref{fig:ratio_dm31}.  This
complexity permits relatively tight
constraints on that parameter.  The
optical efficiency of the DOMs is the leading contributor to the uncertainty budget
for this parameter, and in fact is the systematic which is most
closely correlated individually with the oscillation parameters (see
Fig.~\ref{fig:correlations} below),
 but even so, perfect knowledge of the optical efficiency would
only improve the precision of the measurement of this parameter by
about 20\% given current angular and energy resolutions and current knowledge
of other systematic effects.  

In practice, all eleven systematic uncertainties and both oscillation
parameters are fitted simultaneously.  The optimization routine
explores all possible combinations of systematics that could mimic the more
tightly localized effect of oscillation physics at specific ranges of
energy and zenith angle.  The correlation matrix shown
in Fig.~\ref{fig:correlations} illustrates the correlations between
the systematic parameters and the oscillation parameters.
While some of the systematics are correlated with each other, the wide
ranges of baselines and energies over which we 
observe neutrinos enable us to disentangle these effects from 
the distortions of the observed atmospheric neutrino
flux produced by neutrino oscillations.



%% file: main.bbl
\begin{thebibliography}{47}
\expandafter\ifx\csname natexlab\endcsname\relax\def\natexlab#1{#1}\fi
\expandafter\ifx\csname bibnamefont\endcsname\relax
  \def\bibnamefont#1{#1}\fi
\expandafter\ifx\csname bibfnamefont\endcsname\relax
  \def\bibfnamefont#1{#1}\fi
\expandafter\ifx\csname citenamefont\endcsname\relax
  \def\citenamefont#1{#1}\fi
\expandafter\ifx\csname url\endcsname\relax
  \def\url#1{\texttt{#1}}\fi
\expandafter\ifx\csname urlprefix\endcsname\relax\def\urlprefix{URL }\fi
\providecommand{\bibinfo}[2]{#2}
\providecommand{\eprint}[2][]{\url{#2}}

\bibitem[{\citenamefont{Fukuda et~al.}(1998)}]{Fukuda:1998mi}
\bibinfo{author}{\bibfnamefont{Y.}~\bibnamefont{Fukuda}} \bibnamefont{et~al.}
  (\bibinfo{collaboration}{Super-Kamiokande}), \bibinfo{journal}{Phys. Rev.
  Lett.} \textbf{\bibinfo{volume}{81}}, \bibinfo{pages}{1562}
  (\bibinfo{year}{1998}), \eprint{hep-ex/9807003}.

\bibitem[{\citenamefont{Ahmad et~al.}(2001)}]{Ahmad:2001an}
\bibinfo{author}{\bibfnamefont{Q.~R.} \bibnamefont{Ahmad}} \bibnamefont{et~al.}
  (\bibinfo{collaboration}{SNO}), \bibinfo{journal}{Phys. Rev. Lett.}
  \textbf{\bibinfo{volume}{87}}, \bibinfo{pages}{071301}
  (\bibinfo{year}{2001}), \eprint{nucl-ex/0106015}.

\bibitem[{\citenamefont{Pontecorvo}(1957)}]{Pontecorvo:1957cp}
\bibinfo{author}{\bibfnamefont{B.}~\bibnamefont{Pontecorvo}},
  \bibinfo{journal}{Sov. Phys. JETP} \textbf{\bibinfo{volume}{6}},
  \bibinfo{pages}{429} (\bibinfo{year}{1957}), \bibinfo{note}{[Zh. Eksp. Teor.
  Fiz.33,549(1957)]}.

\bibitem[{\citenamefont{Maki et~al.}(1962)\citenamefont{Maki, Nakagawa, and
  Sakata}}]{Maki:1962mu}
\bibinfo{author}{\bibfnamefont{Z.}~\bibnamefont{Maki}},
  \bibinfo{author}{\bibfnamefont{M.}~\bibnamefont{Nakagawa}}, \bibnamefont{and}
  \bibinfo{author}{\bibfnamefont{S.}~\bibnamefont{Sakata}},
  \bibinfo{journal}{Prog. Theor. Phys.} \textbf{\bibinfo{volume}{28}},
  \bibinfo{pages}{870} (\bibinfo{year}{1962}).

\bibitem[{\citenamefont{Volkova}(1980)}]{Volkova:1980sw}
\bibinfo{author}{\bibfnamefont{L.~V.} \bibnamefont{Volkova}},
  \bibinfo{journal}{Sov. J. Nucl. Phys.} \textbf{\bibinfo{volume}{31}},
  \bibinfo{pages}{784} (\bibinfo{year}{1980}), \bibinfo{note}{[Yad.
  Fiz.31,1510(1980)]}.

\bibitem[{\citenamefont{Barr et~al.}(2004)\citenamefont{Barr, Gaisser, Lipari,
  Robbins, and Stanev}}]{Barr:2004br}
\bibinfo{author}{\bibfnamefont{G.~D.} \bibnamefont{Barr}},
  \bibinfo{author}{\bibfnamefont{T.~K.} \bibnamefont{Gaisser}},
  \bibinfo{author}{\bibfnamefont{P.}~\bibnamefont{Lipari}},
  \bibinfo{author}{\bibfnamefont{S.}~\bibnamefont{Robbins}}, \bibnamefont{and}
  \bibinfo{author}{\bibfnamefont{T.}~\bibnamefont{Stanev}},
  \bibinfo{journal}{Phys. Rev.} \textbf{\bibinfo{volume}{D70}},
  \bibinfo{pages}{023006} (\bibinfo{year}{2004}), \eprint{astro-ph/0403630}.

\bibitem[{\citenamefont{Honda et~al.}(2015)\citenamefont{Honda, Athar, Kajita,
  Kasahara, and Midorikawa}}]{Honda:2015fha}
\bibinfo{author}{\bibfnamefont{M.}~\bibnamefont{Honda}},
  \bibinfo{author}{\bibfnamefont{M.~S.} \bibnamefont{Athar}},
  \bibinfo{author}{\bibfnamefont{T.}~\bibnamefont{Kajita}},
  \bibinfo{author}{\bibfnamefont{K.}~\bibnamefont{Kasahara}}, \bibnamefont{and}
  \bibinfo{author}{\bibfnamefont{S.}~\bibnamefont{Midorikawa}},
  \bibinfo{journal}{Phys. Rev.} \textbf{\bibinfo{volume}{D92}},
  \bibinfo{pages}{023004} (\bibinfo{year}{2015}), \eprint{1502.03916}.

\bibitem[{\citenamefont{Wolfenstein}(1978)}]{Wolfenstein:1977ue}
\bibinfo{author}{\bibfnamefont{L.}~\bibnamefont{Wolfenstein}},
  \bibinfo{journal}{Phys. Rev.} \textbf{\bibinfo{volume}{D17}},
  \bibinfo{pages}{2369} (\bibinfo{year}{1978}).

\bibitem[{\citenamefont{Petcov}(1998)}]{Petcov:1998su}
\bibinfo{author}{\bibfnamefont{S.~T.} \bibnamefont{Petcov}},
  \bibinfo{journal}{Phys. Lett.} \textbf{\bibinfo{volume}{B434}},
  \bibinfo{pages}{321} (\bibinfo{year}{1998}), \eprint{hep-ph/9805262}.

\bibitem[{\citenamefont{Akhmedov et~al.}(2007)\citenamefont{Akhmedov, Maltoni,
  and Smirnov}}]{Akhmedov:2006hb}
\bibinfo{author}{\bibfnamefont{E.~K.} \bibnamefont{Akhmedov}},
  \bibinfo{author}{\bibfnamefont{M.}~\bibnamefont{Maltoni}}, \bibnamefont{and}
  \bibinfo{author}{\bibfnamefont{A.~{\relax Yu}.} \bibnamefont{Smirnov}},
  \bibinfo{journal}{JHEP} \textbf{\bibinfo{volume}{05}}, \bibinfo{pages}{077}
  (\bibinfo{year}{2007}), \eprint{hep-ph/0612285}.

\bibitem[{\citenamefont{Akhmedov et~al.}(2008)\citenamefont{Akhmedov, Maltoni,
  and Smirnov}}]{Akhmedov:2008qt}
\bibinfo{author}{\bibfnamefont{E.~K.} \bibnamefont{Akhmedov}},
  \bibinfo{author}{\bibfnamefont{M.}~\bibnamefont{Maltoni}}, \bibnamefont{and}
  \bibinfo{author}{\bibfnamefont{A.~{\relax Yu}.} \bibnamefont{Smirnov}},
  \bibinfo{journal}{JHEP} \textbf{\bibinfo{volume}{06}}, \bibinfo{pages}{072}
  (\bibinfo{year}{2008}), \eprint{0804.1466}.

\bibitem[{\citenamefont{Adamson et~al.}(2013)}]{Adamson:2013whj}
\bibinfo{author}{\bibfnamefont{P.}~\bibnamefont{Adamson}} \bibnamefont{et~al.}
  (\bibinfo{collaboration}{MINOS}), \bibinfo{journal}{Phys. Rev. Lett.}
  \textbf{\bibinfo{volume}{110}}, \bibinfo{pages}{251801}
  (\bibinfo{year}{2013}), \eprint{1304.6335}.

\bibitem[{\citenamefont{Abe et~al.}(2017)}]{Abe:2017uxa}
\bibinfo{author}{\bibfnamefont{K.}~\bibnamefont{Abe}} \bibnamefont{et~al.}
  (\bibinfo{collaboration}{T2K}), \bibinfo{journal}{Phys. Rev. Lett.}
  \textbf{\bibinfo{volume}{118}}, \bibinfo{pages}{151801}
  (\bibinfo{year}{2017}), \eprint{1701.00432}.

\bibitem[{\citenamefont{Adamson et~al.}(2017)}]{Adamson:2017qqn}
\bibinfo{author}{\bibfnamefont{P.}~\bibnamefont{Adamson}} \bibnamefont{et~al.}
  (\bibinfo{collaboration}{NOvA}), \bibinfo{journal}{Phys. Rev. Lett.}
  \textbf{\bibinfo{volume}{118}}, \bibinfo{pages}{151802}
  (\bibinfo{year}{2017}), \eprint{1701.05891}.

\bibitem[{\citenamefont{An et~al.}(2017)}]{An:2016ses}
\bibinfo{author}{\bibfnamefont{F.~P.} \bibnamefont{An}} \bibnamefont{et~al.}
  (\bibinfo{collaboration}{Daya Bay}), \bibinfo{journal}{Phys. Rev.}
  \textbf{\bibinfo{volume}{D95}}, \bibinfo{pages}{072006}
  (\bibinfo{year}{2017}), \eprint{1610.04802}.

\bibitem[{\citenamefont{Wendell}(2015)}]{Wendell:2014dka}
\bibinfo{author}{\bibfnamefont{R.}~\bibnamefont{Wendell}}
  (\bibinfo{collaboration}{Super-Kamiokande}), \bibinfo{journal}{AIP Conf.
  Proc.} \textbf{\bibinfo{volume}{1666}}, \bibinfo{pages}{100001}
  (\bibinfo{year}{2015}), \eprint{1412.5234}.

\bibitem[{\citenamefont{Formaggio and Zeller}(2012)}]{Formaggio:2013kya}
\bibinfo{author}{\bibfnamefont{J.~A.} \bibnamefont{Formaggio}}
  \bibnamefont{and} \bibinfo{author}{\bibfnamefont{G.~P.}
  \bibnamefont{Zeller}}, \bibinfo{journal}{Rev. Mod. Phys.}
  \textbf{\bibinfo{volume}{84}}, \bibinfo{pages}{1307} (\bibinfo{year}{2012}),
  \eprint{1305.7513}.

\bibitem[{\citenamefont{Friedland et~al.}(2004)\citenamefont{Friedland,
  Lunardini, and Maltoni}}]{Friedland:2004ah}
\bibinfo{author}{\bibfnamefont{A.}~\bibnamefont{Friedland}},
  \bibinfo{author}{\bibfnamefont{C.}~\bibnamefont{Lunardini}},
  \bibnamefont{and} \bibinfo{author}{\bibfnamefont{M.}~\bibnamefont{Maltoni}},
  \bibinfo{journal}{Phys. Rev.} \textbf{\bibinfo{volume}{D70}},
  \bibinfo{pages}{111301} (\bibinfo{year}{2004}), \eprint{hep-ph/0408264}.

\bibitem[{\citenamefont{Friedland and Lunardini}(2005)}]{Friedland:2005vy}
\bibinfo{author}{\bibfnamefont{A.}~\bibnamefont{Friedland}} \bibnamefont{and}
  \bibinfo{author}{\bibfnamefont{C.}~\bibnamefont{Lunardini}},
  \bibinfo{journal}{Phys. Rev.} \textbf{\bibinfo{volume}{D72}},
  \bibinfo{pages}{053009} (\bibinfo{year}{2005}), \eprint{hep-ph/0506143}.

\bibitem[{\citenamefont{Ohlsson et~al.}(2013)\citenamefont{Ohlsson, Zhang, and
  Zhou}}]{Ohlsson:2013epa}
\bibinfo{author}{\bibfnamefont{T.}~\bibnamefont{Ohlsson}},
  \bibinfo{author}{\bibfnamefont{H.}~\bibnamefont{Zhang}}, \bibnamefont{and}
  \bibinfo{author}{\bibfnamefont{S.}~\bibnamefont{Zhou}},
  \bibinfo{journal}{Phys. Rev.} \textbf{\bibinfo{volume}{D88}},
  \bibinfo{pages}{013001} (\bibinfo{year}{2013}), \eprint{1303.6130}.

\bibitem[{\citenamefont{Esmaili and Smirnov}(2013)}]{Esmaili:2013fva}
\bibinfo{author}{\bibfnamefont{A.}~\bibnamefont{Esmaili}} \bibnamefont{and}
  \bibinfo{author}{\bibfnamefont{A.~{\relax Yu}.} \bibnamefont{Smirnov}},
  \bibinfo{journal}{JHEP} \textbf{\bibinfo{volume}{06}}, \bibinfo{pages}{026}
  (\bibinfo{year}{2013}), \eprint{1304.1042}.

\bibitem[{\citenamefont{Gonzalez-Garcia and
  Maltoni}(2013)}]{Gonzalez-Garcia:2013usa}
\bibinfo{author}{\bibfnamefont{M.~C.} \bibnamefont{Gonzalez-Garcia}}
  \bibnamefont{and} \bibinfo{author}{\bibfnamefont{M.}~\bibnamefont{Maltoni}},
  \bibinfo{journal}{JHEP} \textbf{\bibinfo{volume}{09}}, \bibinfo{pages}{152}
  (\bibinfo{year}{2013}), \eprint{1307.3092}.

\bibitem[{\citenamefont{Mocioiu and Wright}(2015)}]{Mocioiu:2014gua}
\bibinfo{author}{\bibfnamefont{I.}~\bibnamefont{Mocioiu}} \bibnamefont{and}
  \bibinfo{author}{\bibfnamefont{W.}~\bibnamefont{Wright}},
  \bibinfo{journal}{Nucl. Phys.} \textbf{\bibinfo{volume}{B893}},
  \bibinfo{pages}{376} (\bibinfo{year}{2015}), \eprint{1410.6193}.

\bibitem[{\citenamefont{Choubey and Ohlsson}(2014)}]{Choubey:2014iia}
\bibinfo{author}{\bibfnamefont{S.}~\bibnamefont{Choubey}} \bibnamefont{and}
  \bibinfo{author}{\bibfnamefont{T.}~\bibnamefont{Ohlsson}},
  \bibinfo{journal}{Phys. Lett.} \textbf{\bibinfo{volume}{B739}},
  \bibinfo{pages}{357} (\bibinfo{year}{2014}), \eprint{1410.0410}.

\bibitem[{\citenamefont{Coloma and Schwetz}(2016)}]{Coloma:2016gei}
\bibinfo{author}{\bibfnamefont{P.}~\bibnamefont{Coloma}} \bibnamefont{and}
  \bibinfo{author}{\bibfnamefont{T.}~\bibnamefont{Schwetz}},
  \bibinfo{journal}{Phys. Rev.} \textbf{\bibinfo{volume}{D94}},
  \bibinfo{pages}{055005} (\bibinfo{year}{2016}), \bibinfo{note}{[Erratum:
  Phys. Rev. D95, 079903 (2017)]}, \eprint{1604.05772}.

\bibitem[{\citenamefont{Liao et~al.}(2016)\citenamefont{Liao, Marfatia, and
  Whisnant}}]{Liao:2016hsa}
\bibinfo{author}{\bibfnamefont{J.}~\bibnamefont{Liao}},
  \bibinfo{author}{\bibfnamefont{D.}~\bibnamefont{Marfatia}}, \bibnamefont{and}
  \bibinfo{author}{\bibfnamefont{K.}~\bibnamefont{Whisnant}},
  \bibinfo{journal}{Phys. Rev.} \textbf{\bibinfo{volume}{D93}},
  \bibinfo{pages}{093016} (\bibinfo{year}{2016}), \eprint{1601.00927}.

\bibitem[{\citenamefont{Aartsen et~al.}(2017{\natexlab{a}})}]{Aartsen:2017bap}
\bibinfo{author}{\bibfnamefont{M.~G.} \bibnamefont{Aartsen}}
  \bibnamefont{et~al.} (\bibinfo{collaboration}{IceCube}),
  \bibinfo{journal}{Phys. Rev.} \textbf{\bibinfo{volume}{D95}},
  \bibinfo{pages}{112002} (\bibinfo{year}{2017}{\natexlab{a}}),
  \eprint{1702.05160}.

\bibitem[{\citenamefont{Aartsen et~al.}(2015)}]{Aartsen:2014yll}
\bibinfo{author}{\bibfnamefont{M.~G.} \bibnamefont{Aartsen}}
  \bibnamefont{et~al.} (\bibinfo{collaboration}{IceCube}),
  \bibinfo{journal}{Phys. Rev.} \textbf{\bibinfo{volume}{D91}},
  \bibinfo{pages}{072004} (\bibinfo{year}{2015}), \eprint{1410.7227}.

\bibitem[{\citenamefont{Aartsen et~al.}(2017{\natexlab{b}})}]{Aartsen:2016nxy}
\bibinfo{author}{\bibfnamefont{M.~G.} \bibnamefont{Aartsen}}
  \bibnamefont{et~al.} (\bibinfo{collaboration}{IceCube}),
  \bibinfo{journal}{JINST} \textbf{\bibinfo{volume}{12}},
  \bibinfo{pages}{P03012} (\bibinfo{year}{2017}{\natexlab{b}}),
  \eprint{1612.05093}.

\bibitem[{\citenamefont{Abbasi et~al.}(2009)}]{Abbasi:2008aa}
\bibinfo{author}{\bibfnamefont{R.}~\bibnamefont{Abbasi}} \bibnamefont{et~al.}
  (\bibinfo{collaboration}{IceCube}), \bibinfo{journal}{Nucl. Instrum. Meth.}
  \textbf{\bibinfo{volume}{A601}}, \bibinfo{pages}{294} (\bibinfo{year}{2009}),
  \eprint{0810.4930}.

\bibitem[{\citenamefont{Abbasi et~al.}(2010)}]{Abbasi:2010vc}
\bibinfo{author}{\bibfnamefont{R.}~\bibnamefont{Abbasi}} \bibnamefont{et~al.}
  (\bibinfo{collaboration}{IceCube}), \bibinfo{journal}{Nucl. Instrum. Meth.}
  \textbf{\bibinfo{volume}{A618}}, \bibinfo{pages}{139} (\bibinfo{year}{2010}),
  \eprint{1002.2442}.

\bibitem[{\citenamefont{Abbasi et~al.}(2012)}]{Collaboration:2011ym}
\bibinfo{author}{\bibfnamefont{R.}~\bibnamefont{Abbasi}} \bibnamefont{et~al.}
  (\bibinfo{collaboration}{IceCube}), \bibinfo{journal}{Astropart. Phys.}
  \textbf{\bibinfo{volume}{35}}, \bibinfo{pages}{615} (\bibinfo{year}{2012}),
  \eprint{1109.6096}.

\bibitem[{\citenamefont{Andreopoulos et~al.}(2010)}]{Andreopoulos:2009rq}
\bibinfo{author}{\bibfnamefont{C.}~\bibnamefont{Andreopoulos}}
  \bibnamefont{et~al.}, \bibinfo{journal}{Nucl. Instrum. Meth.}
  \textbf{\bibinfo{volume}{A614}}, \bibinfo{pages}{87} (\bibinfo{year}{2010}),
  \eprint{0905.2517}.

\bibitem[{\citenamefont{Agostinelli et~al.}(2003)}]{Agostinelli:2002hh}
\bibinfo{author}{\bibfnamefont{S.}~\bibnamefont{Agostinelli}}
  \bibnamefont{et~al.} (\bibinfo{collaboration}{GEANT4}),
  \bibinfo{journal}{Nucl. Instrum. Meth.} \textbf{\bibinfo{volume}{A506}},
  \bibinfo{pages}{250} (\bibinfo{year}{2003}).

\bibitem[{\citenamefont{Radel and Wiebusch}(2012)}]{Radel:2012kw}
\bibinfo{author}{\bibfnamefont{L.}~\bibnamefont{Radel}} \bibnamefont{and}
  \bibinfo{author}{\bibfnamefont{C.}~\bibnamefont{Wiebusch}},
  \bibinfo{journal}{Astropart. Phys.} \textbf{\bibinfo{volume}{38}},
  \bibinfo{pages}{53} (\bibinfo{year}{2012}), \eprint{1206.5530}.

\bibitem[{\citenamefont{Koehne et~al.}(2013)\citenamefont{Koehne, Frantzen,
  Schmitz, Fuchs, Rhode, Chirkin, and Becker~Tjus}}]{Koehne:2013gpa}
\bibinfo{author}{\bibfnamefont{J.~H.} \bibnamefont{Koehne}},
  \bibinfo{author}{\bibfnamefont{K.}~\bibnamefont{Frantzen}},
  \bibinfo{author}{\bibfnamefont{M.}~\bibnamefont{Schmitz}},
  \bibinfo{author}{\bibfnamefont{T.}~\bibnamefont{Fuchs}},
  \bibinfo{author}{\bibfnamefont{W.}~\bibnamefont{Rhode}},
  \bibinfo{author}{\bibfnamefont{D.}~\bibnamefont{Chirkin}}, \bibnamefont{and}
  \bibinfo{author}{\bibfnamefont{J.}~\bibnamefont{Becker~Tjus}},
  \bibinfo{journal}{Comput. Phys. Commun.} \textbf{\bibinfo{volume}{184}},
  \bibinfo{pages}{2070} (\bibinfo{year}{2013}).

\bibitem[{\citenamefont{Kopper et~al.}()}]{CLSIM}
\bibinfo{author}{\bibfnamefont{C.}~\bibnamefont{Kopper}} \bibnamefont{et~al.},
  \bibinfo{note}{\url{https://github.com/claudiok/clsim}}.

\bibitem[{\citenamefont{Aartsen et~al.}(2013)}]{Aartsen:2013rt}
\bibinfo{author}{\bibfnamefont{M.~G.} \bibnamefont{Aartsen}}
  \bibnamefont{et~al.} (\bibinfo{collaboration}{IceCube}),
  \bibinfo{journal}{Nucl. Instrum. Meth.} \textbf{\bibinfo{volume}{A711}},
  \bibinfo{pages}{73} (\bibinfo{year}{2013}), \eprint{1301.5361}.

\bibitem[{\citenamefont{Feroz et~al.}(2009)\citenamefont{Feroz, Hobson, and
  Bridges}}]{Feroz:2008xx}
\bibinfo{author}{\bibfnamefont{F.}~\bibnamefont{Feroz}},
  \bibinfo{author}{\bibfnamefont{M.~P.} \bibnamefont{Hobson}},
  \bibnamefont{and} \bibinfo{author}{\bibfnamefont{M.}~\bibnamefont{Bridges}},
  \bibinfo{journal}{Mon. Not. Roy. Astron. Soc.}
  \textbf{\bibinfo{volume}{398}}, \bibinfo{pages}{1601} (\bibinfo{year}{2009}),
  \eprint{0809.3437}.

\bibitem[{\citenamefont{Hocker et~al.}(2007)}]{Hocker:2007ht}
\bibinfo{author}{\bibfnamefont{A.}~\bibnamefont{Hocker}} \bibnamefont{et~al.},
  \bibinfo{journal}{PoS} \textbf{\bibinfo{volume}{ACAT}}, \bibinfo{pages}{040}
  (\bibinfo{year}{2007}), \eprint{physics/0703039}.

\bibitem[{\citenamefont{Ahrens et~al.}(2004)}]{Ahrens:2003fg}
\bibinfo{author}{\bibfnamefont{J.}~\bibnamefont{Ahrens}} \bibnamefont{et~al.}
  (\bibinfo{collaboration}{AMANDA}), \bibinfo{journal}{Nucl. Instrum. Meth.}
  \textbf{\bibinfo{volume}{A524}}, \bibinfo{pages}{169} (\bibinfo{year}{2004}),
  \eprint{astro-ph/0407044}.

\bibitem[{\citenamefont{James and Roos}(1975)}]{James:1975dr}
\bibinfo{author}{\bibfnamefont{F.}~\bibnamefont{James}} \bibnamefont{and}
  \bibinfo{author}{\bibfnamefont{M.}~\bibnamefont{Roos}},
  \bibinfo{journal}{Comput. Phys. Commun.} \textbf{\bibinfo{volume}{10}},
  \bibinfo{pages}{343} (\bibinfo{year}{1975}).

\bibitem[{\citenamefont{Wendell et~al.}()}]{prob3pp}
\bibinfo{author}{\bibfnamefont{R.}~\bibnamefont{Wendell}} \bibnamefont{et~al.},
  \bibinfo{note}{\url{http://www.phy.duke.edu/~raw22/public/Prob3++}}.

\bibitem[{\citenamefont{Bodek and Yang}(2003)}]{Bodek:2002ps}
\bibinfo{author}{\bibfnamefont{A.}~\bibnamefont{Bodek}} \bibnamefont{and}
  \bibinfo{author}{\bibfnamefont{U.~K.} \bibnamefont{Yang}},
  \bibinfo{journal}{J. Phys.} \textbf{\bibinfo{volume}{G29}},
  \bibinfo{pages}{1899} (\bibinfo{year}{2003}), \eprint{hep-ex/0210024}.

\bibitem[{\citenamefont{Katori et~al.}(2016)\citenamefont{Katori, Lasorak,
  Mandalia, and Terri}}]{Katori:2016blu}
\bibinfo{author}{\bibfnamefont{T.}~\bibnamefont{Katori}},
  \bibinfo{author}{\bibfnamefont{P.}~\bibnamefont{Lasorak}},
  \bibinfo{author}{\bibfnamefont{S.}~\bibnamefont{Mandalia}}, \bibnamefont{and}
  \bibinfo{author}{\bibfnamefont{R.}~\bibnamefont{Terri}},
  \bibinfo{journal}{JPS Conf. Proc.} \textbf{\bibinfo{volume}{12}},
  \bibinfo{pages}{010033} (\bibinfo{year}{2016}), \eprint{1602.00083}.

\bibitem[{\citenamefont{Barr et~al.}(2006)\citenamefont{Barr, Robbins, Gaisser,
  and Stanev}}]{Barr:2006it}
\bibinfo{author}{\bibfnamefont{G.~D.} \bibnamefont{Barr}},
  \bibinfo{author}{\bibfnamefont{S.}~\bibnamefont{Robbins}},
  \bibinfo{author}{\bibfnamefont{T.~K.} \bibnamefont{Gaisser}},
  \bibnamefont{and} \bibinfo{author}{\bibfnamefont{T.}~\bibnamefont{Stanev}},
  \bibinfo{journal}{Phys. Rev.} \textbf{\bibinfo{volume}{D74}},
  \bibinfo{pages}{094009} (\bibinfo{year}{2006}), \eprint{astro-ph/0611266}.

\bibitem[{\citenamefont{Feldman and Cousins}(1998)}]{Feldman:1997qc}
\bibinfo{author}{\bibfnamefont{G.~J.} \bibnamefont{Feldman}} \bibnamefont{and}
  \bibinfo{author}{\bibfnamefont{R.~D.} \bibnamefont{Cousins}},
  \bibinfo{journal}{Phys. Rev.} \textbf{\bibinfo{volume}{D57}},
  \bibinfo{pages}{3873} (\bibinfo{year}{1998}), \eprint{physics/9711021}.

\end{thebibliography}
